\documentclass[12pt,preprint]{aastex}
\def\pc{{\rm\,pc}}
\def\yrs{{\rm\,yrs}}

\def\Myr{{\rm\,Myr}}

\def\AU{{\rm\,AU}}
\def\o{{\rm\,o}}
\def\days{{\rm\,days}}

\def\erf{{\rm\,erf}}
\newcommand{\dgrs}{\ensuremath{^\circ}}

\shortauthors{Koch \& Hansen}
\shorttitle{Black Hole Dynamics}

\begin{document}
% \twocolumn 
\title{Mergers of Black Holes in the Galactic Center} 
 
\author{F. Elliott Koch and Bradley M.S. Hansen} 
\affil{Department of Physics and Astronomy, University of California, 
    Los Angeles, CA} 

\begin{abstract} 
We present the results of three~body simulations focused on understanding 
the fates of intermediate mass black holes (IBH) that drift within the 
central 0.5 pc of the Galaxy. In particular, we modeled the interactions 
between pairs of $4000 {\rm M}_{\odot}$ black holes as they orbit a central black
hole of mass $4 \times 10^6 {\rm M}_{\odot}$.
The simulations performed assume a  
Schwarzschild geometry and account for Chandrasekhar dynamical friction as  
well as acceleration resulting from energy lost due to gravitational radiation. 
We found the branching ratio for one of the orbiting IBHs to merge with 
the CBH was 0.95 and is independent of the inner IBH's initial eccentricity 
as well as the rate of sinking.  This, coupled with an infall rate of $\sim 10^7$ yrs 
for an IBH to drift into the Galactic center, results in an IBH-CBH  
merger every $\lesssim 11$ Myrs.  Lastly we found that the IBH-IBH-CBH 
triple body system ``resets'' itself, in the sense that a system with an inner IBH 
with an initially circular orbit generally left behind an IBH with a  
large eccentricity, whereas a system in which the 
inner IBH had a high eccentricity ($e_0 \sim 0.9$) usually left a 
remnant with low eccentricity. Branching ratios for different outcomes are also
similar in the two cases.
%Additionally, interactions between the  
%two IBHs cause the black holes to have periodically develop oribits with 
%large angles of inclination (at times retrograde) with respect to the 
%Galactic plane, which could explain the random distribution of one group 
%of stars in the Galactic center.   

\end{abstract}  
  
\section{Introduction}  
\label{sec:Intro}  
  
Starting with the discovery of quasars \citep{Schmidt63, Greenstein64, Greenstein64b},  
 we have gradually come to  
appreciate the existence of massive, dark objects in the centers of  
almost all galaxies \citep{Bell69, Richstone98}, although  
their origins are still unclear. The nature of such a central 
dark object is best  
constrained in the center of our own Galaxy. 
Over the last decade, the infra-red   
monitoring of the Galactic center has led to severe 
dynamical constraints on the  
nature of the Milky Way's central object 
\citep{Eckart97, Ghez98, Schodel02, Ghez05},   
 leaving little doubt that it is a black hole with mass   
$3.7 \pm 0.2 \times 10^6 {\rm M}_{\odot}$.  
  
An equally interesting result to emerge from these studies 
is that the stellar population  
within the central parsec of the galaxy contains an 
unexpected component of   
young, massive stars 
\citep{Sanders92, Morris93, Genzel97, Ghez03, Paumard06}.   
The origin of these stars is a question of interest because the strong tidal  
fields close to the central black hole (CBH) are large enough to shear apart 
a normal  
molecular cloud at the locations where the stars are seen \citep{Morris93}. 
Two principal  
classes of alternative models have been investigated -- one which 
postulates that the  
stars form from the gravitational instability of a quasar-like 
accretion disk around  
the CBH during an earlier period of high nuclear activity 
\citep{Levin03} and  
 another which proposes that the stars migrate   
inwards as part of a stellar cluster which sinks   
towards the CBH by dynamical friction and is eventually broken 
up by the strong tidal  
field \citep{Gerhard01}. A recent addition to this latter scenario 
is the possibility that such  
a cluster contains an intermediate mass black hole \citep{Hansen03}, which  
serves to bind the cluster tighter and slow the internal 
dynamical relaxation, allowing  
the cluster to survive long enough to deposit the stars in their 
observed locations.  
  
Each of these scenarios have their proponents, and 
the arguments for and against  
are presented elsewhere (e.g. \citep{Paumard06, Berukoff06}). The goal  
of this paper is to examine the implications the cluster infall 
scenario for the   
merger history of the central black hole. In particular, once the 
intermediate mass  
black hole (IBH) is stripped of its coterie of cluster stars, it 
continues to sink towards  
the center due to dynamical friction. We wish to examine the 
final fate of such an IBH and how that fate is reached.  
We will explore in \S\ref{sec:Merge} reasons to believe that the rate of   
infall of such IBH  
into the central parsec  
may actually be larger than the rate at which an individual 
IBH would merge with the CBH.  
This then opens the possibility of multiple IBH in orbit, so that in 
\S\ref{sec:Orbs} we  
examine the gravitational interactions of such CBH-IBH-IBH 
triples and explore the  
probabilities of various outcomes. The results of this 
study have a variety of interesting  
applications, not the least of which is a prediction for 
the kinds of signals we might  
expect to observe with future space based gravitational wave observatories.   
We discuss this  
further in \S\ref{sec:Discuss}.  
  
\section{The fate of a sinking IBH}  
\label{sec:Merge}  
  
Let us consider an IBH, surrounded by a handful of stars,   
the last remnants of the parent cluster, as it sinks towards the CBH.   
Full dynamical simulations of this scenario can be  
found in \citet{Berukoff06} but, for our purposes, we 
consider the simplified model  
in which the remaining stars have relaxed sufficiently to be   
distributed in a Bahcall-Wolf  
density profile ($\propto r^{-7/4}$).   
We normalise our cluster remnant to contain 50 stars  
within the Roche lobe (for a 4000${\rm M}_{\odot}$ IBH   
and a CBH mass $4 \times 10^6 {\rm M}_{\odot}$)  
when located at 1 pc. As the cluster sinks,   
the Roche lobe shrinks, and stars are stripped  
and deposited into orbits about the CBH. The last star in this scenario is stripped at  
separations  
\begin{equation}  
 R_{strip} \sim 0.04 pc \left( \frac{\mu}{0.001} \right)^{-1/3}  
\end{equation}  
which correspond to angular separations $\sim 1''$. Interior to this radius nearly the  
entire cluster is stripped, although the full simulations suggest a single closely bound  
star may be retained. After this, the IBH will sink as an independent, and usually  
unobservable, entity.  
  
The IBH sinks by dynamical friction -- that is, losing energy by scattering stars  
in the local density field. The sinking will therefore slow when the IBH has sunk far  
enough that the mass in the potential reservoir of local stars is small compared that   
of the IBH \citep{Begelman80}.  
 At this point, the IBH can eject all the available stellar mass without significantly  
changing it's binding energy. Thus, we expect the central region to be largely evacuated  
of stars, refilled only by dynamical processes that scatter stars inwards from larger radii.  
The observed density profile of the Galactic center \citep{Genzel03} is  
\begin{equation}  
 \rho (R) = 1.2 \times 10^6 {\rm M}_{\odot}.pc^{-3} \left( \frac{R}{0.4 pc} \right)^{-1.4}  
\end{equation}  
at distances interior to $R=0.4 pc$. With this profile and our nominal IBH mass of  
4000 ${\rm M}_{\odot}$, the `stalling radius'$R_{stall}$, where the IBH mass equals the  
 mass in field  
stars interior to that radius, lies at 0.02~pc. Thus, in the simplest version of this  
scenario, the dynamical drag that drives the IBH inspiral should tail off on scales  
$\sim 0.02~$pc, i.e. this is where, in the absence of other effects, the IBH should  
stall. Such behaviour is, in fact, seen in full N-body simulations of the inspiral
process \citep{MME, LockBaum}.
  
Beyond this point, the further evolution of the IBH-CBH binary should follow the  
kind of pattern outlined by \citet{Begelman80} for supermassive  
black hole binaries in galactic nuclei, albeit with a somewhat more extreme mass  
ratio. The timescale for merger by gravitational wave emission (assuming a circular  
orbit) from this distance is  
%\begin{equation}  
\begin{eqnarray}  
  T_{GW} & \sim & 10^{14} years \left( \frac{R_{stall}}{0.01 pc} \right)^4  
    \left( \frac{M_{IBH}}{4000 {\rm M}_{\odot}} \right)^{-1}  \\  
   & \times & \left( \frac{M_{CBH}}{4 \times 10^6 {\rm M}_{\odot}} \right)^{-2}.  
  \nonumber
  \label{eq:Tgw}  
\end{eqnarray}  
%\end{equation}  
On shorter timescales, namely that of the field two-body relaxation time, the stellar  
loss cone is refilled and the IBH will lose further binding energy in ejecting those  
stars which cross its orbit. A detailed description of this process is not necessary  
for our purposes, since the field relaxation time is $\sim 10^9$ years in the central  
Galactic parsec, so that the inspiral time is bounded from below by such a timescale.  
  
The reason the long merger times are of interest is because there are several reasons  
to believe that the interval between IBH deliveries to the central parsec may be  
considerably shorter than $10^9$~years. Perhaps the simplest reason is that, if we  
take the age of the young massive stars ($\sim 10^7$~years) as a characteristic  
timescale, then we are faced with the choice between assuming that processes like  
cluster infall occur regularly on such timescales or assigning a very special  
and un-Copernican nature  
to our current epoch. Furthermore, the notion that stellar clusters might contain  
IBH finds support in the simulations of cluster formation by \citet{Zwart99}  
 and \citet{Freitag05}. If the cluster is dense enough, then  
mass segregation brings massive main sequence stars to the center before they finish  
their stellar evolution, resulting in an epoch of stellar collisions and mergers. The  
upshot of this process is the formation of a massive, stellar object a fraction   
$\sim 10^{-3}$  
of the total cluster mass. The subsequent evolution of such an object is far from clear,  
but unless it manages to lose a considerable fraction of its mass in the form of winds,  
it is likely to, in the end, form a black hole, since it is far too massive to form either  
a white dwarf or neutron star.  
  
Of more direct interest to this paper is the calculation of \citet{Zwart06},  
in which they simulate a population of stellar clusters forming in the central  
100 parsecs of the Galaxy. They find that $\sim 10\%$ of the clusters born in their  
simulation will form such IBH progenitors. The result is that IBH sink towards the center  
with some frequency. Indeed they estimate that the inner 10 pc contains $\sim 50$~IMBH  
at any given time and that IBH enter the central parsec every $\sim 10^7$~years.  
They note that such a high frequency supports the notion \citep{Ebisuzaki01} that  
the CBH may have actually been constructed over a Hubble time via a series of mergers  
between infalling IBH and a central, growing remnant of such mergers. They also discuss  
an order-of-magnitude estimate of the merger dynamics, which is the central focus  
of this paper. We will make a detailed    
comparison in \S~\ref{sec:Discuss}.  
  
Thus, the scenario we intend to investigate is that in which IBH arrive in the  
central parsec at intervals $\sim 10^7$ years, but then stall at radii $\sim 0.02$~pc.  
As a result, we need to investigate the dynamics of two IBH interacting in the   
neighbourhood  
of the CBH.  
  
\section{Triple Systems}  
\label{sec:Orbs}  
Assuming that the orbit of an IBH will decay until there is no longer enough  
stellar mass to affect its orbit, we chose to simulate a three body   
system consisting of an IBH on a circular orbit $\sim 0.02$ pc from the CBH   
and a second IBH that had started its infall sometime later.  
The second IBH is   
introduced at a distance of $\sim 0.05$ pc and a dynamical friction force  
is applied to simulate the orbital decay just as the orbit of  
the inner IBH was assumed to have done.  
Assuming that these two events are uncorrelated, simulations were  
repeated for initial orbital plane separations for the two IBHs between  
$0^{\o}$ and $180^{\o}$ varying in $5^{\o}$ intervals to account for the entire   
orbital sphere.  
Therefore without any loss of generality,   
the initial angle of inclination for the inner  
IBH (one closest to the CBH) was defined to be zero and the outer IBH was   
placed randomly on a circular orbit with a semi-major axis of   
about $\sim 0.05$ pc in its own  
orbital plane.  A complete statistical ensemble was created by repeating  
this for 500 relative phases at each angle of inclination   
between 0 and 180 degrees on 5 degree intervals.     
  
\subsection{Integration Algorithm}  
\label{sec:Int}  
Integration of the orbits were performed using a Burlisch-Stoer integration   
algorithm contained in the Mercury6 program written by \citet{Chambers99}.  
Although the Mercury6 program contains a symplectic   
integration routines that operate more efficiently and  
require less overall computational time, the Burlisch-Stoer   
algorithm more accurately simulates ``close encounters''   
(cases where two masses scatter off of each other) as well   
as highly eccentric orbits, both of   
which occur regularly  
in the results obtained here.   
The tolerance used to determine each Burlisch-Stoer time   
step for the integration was   
set to be $\epsilon = 10^{-15}$ \citep{NumRecipes},  
which was necessary to reduce computational   
errors caused by the highly elliptical  
orbits of the IBH on orbits that would graze the CBH,   
yet still lead to acceptable computing times.  
The Mercury6 program provided an ideal platform for modeling this  
system because its primary mode of operation is for simulating the  
orbital evolution of stellar and planetary systems, and thus the
code is optimised to treat the dynamics in the case where the mass
is dominated by a single central object. 
Rescaling the program so that ${\rm M}_{\odot} \rightarrow 10^6 {\rm M}_{\odot}$ and  
$1\AU \rightarrow 100 \AU$, the CBH would mimic the central star and the orbiting   
IBHs would mimic planets.    
Another useful feature of Mercury6 is that it allows the user to specify  
a ``close encounter'' parameter that enables us to monitor scattering events  
and record the distance of closest   
approach during IBH-IBH interactions.  If the separation of the two black holes  
was ever smaller than three Hill radii, $a(m_{IBH}/m_{CBH})^{1/3}$,   
the program would record the time that  
it occurred and the resulting distance of closest approach.  
Lastly, the Mercury6 program easily enables the program user to define   
accelerations due to non-gravitational forces, in this case forces due to   
dynamical friction and general relativity (see \S~\ref{sec:DynFric} and   
\S~\ref{sec:GR}).  

\subsection{Dynamical Friction}  
\label{sec:DynFric}  
In \S~\ref{sec:Merge}, we discussed how the orbit of an IBH would decay to an orbital  
radius where the IBH could no longer eject enough mass to affect the binding   
energy of its orbit.  The dynamical friction force was   
assumed to be a Chandrasekhar drag force acting on all IBH beyond a radius of  
$R_{stall} \sim 0.02 \pc$ from the CBH.  
%\begin{equation}  
\begin{eqnarray}  
  \frac{d \vec{v}}{dt} & = &
   - 4 \pi \ln \Lambda G^2 \rho M_{IBH} \\  
   & \times & 
     \left[ \erf \left(X\right) - \frac{2X}{\sqrt{\pi}}e^{-X^2} \right]  
     \frac{\vec{v}}{v^3}  
  \nonumber
  \label{eq:drag1}  
 \end{eqnarray}  
%\end{equation}  
Where $X \equiv v/ \sqrt{2} \sigma$, $\ln \Lambda$ is the Coulomb logarithm,  
$\rho$ is the background density and $\sigma$ the velocity dispersion  
\citep{Binney87}.  
Since $Xe^{-X^2} \rightarrow 0$ and $\erf \left(X\right) \rightarrow 1$  
as $X \rightarrow \infty$ and assuming that $\rho$ and $\ln \Lambda$  
change little compared to $\vec{v} / v^3$ for $R > R_{stall}$ on the time  
scales were are interested in, then  
equation (\ref{eq:drag1}) can be rewritten as  
\begin{equation}  
 \frac{d \vec{v}}{dt} = -k \frac{ \vec{v}}{v^3}.  
 \label{eq:drag2}  
\end{equation}  
Therefore assuming a density profile similar to that presented by  
\citet{Genzel03}   
and to ensure that 
$t_{fric} \sim \langle v / \dot{v} \rangle \sim 10^7$ yrs, we set
$k \sim \rho M_{IBH} \ln \lambda \sim 10^{-18} \AU^3 / \days^4$. 
Although $t_{fric}$ is small compared to the relaxation time for the Galactic  
center, a $t_{fric} \sim 10^7 \yrs$ was chosen to be consistent with the  
ages of the young massive stars in the Galactic center   
(in \S\ref{sec:Discuss} we briefly discuss results for $t_{fric} \sim 10^6$ yrs and  
$t_{fric} \sim 10^8$ yrs and note that the final results are well within the 
measurement error associated with the ensemble size for these simulations). 
In order to reduce numerical errors associated with crossing into (or out of) 
the ``loss cone'', equation (\ref{eq:drag2}) was mulitplied by:
\begin{equation}  
 \kappa =  \left\{
  \begin{array}{ll}
   0                      & \mbox{if} y \leq 0, \\  
   \frac{y^2}{2y^2-2y+1}  & \mbox{if} 1 > y > 0, \\  
   1                      & \mbox{otherwise}
 \end{array} \right. 
 \label{eq:scale}  
\end{equation}  
where,  
\begin{displaymath}  
 y = 4 \frac{r - R_{stall}}{R_{stall}}  
\end{displaymath}  
  
\subsection{General Relativity}  
\label{sec:GR}  

Initially our simulations were performed without accounting for any
relativistic effects, with the assumption that an ejection from the 
Galactic center would be the most probable outcome.  We very quickly
realized that this was indeed not the case and that relativistic effects
must be accounted for.  Since we are only interested in the whether or
not an IBH merges with the CBH we chose to neglect Post-Newtonian 
expansions $\lesssim \mathcal{O}(c^{-6})$ which presumably account for
plunge and coalesence effects \citep{Blanchet2003}.  Separating the
contributions to the acceleration from the different expansions scaled
by multiples of the speed of light, the $i^{th}$ component of the 
acceleration due to gravity becomes,
\begin{equation}
	\frac{d v^i}{dt} = a_0^i + \frac{1}{c^2} a_2^i + \frac{1}{c^4} a_4^i
		 + \frac{1}{c^5} a_5^i
	\label{eq:SimpPN}
\end{equation}
where $\vec{a}_2$ is the $1 \mathcal{PN}$, $\vec{a}_4$ is the $2 \mathcal{PN}$ 
and $\vec{a}_5$ is the $2.5 \mathcal{PN}$ \citep{Kupi2006}.  Although
there is extensive literature on Post-Newtonian expansions we chose
to use the applicable perturbations presented in \citet{Damour1981} 
and the references cited within (more recent derivations can be found in
\citet{Itoh2001} and \citet{Blanchet2003})
altered so that the larger of the two masses (the CBH) is defined
as the origin of the coordinate system, as in the Mercury code.  Using similar arguments 
presented in \citet{Gultekin2006}, Post-Newtonian corrections applied
are:
\begin{mathletters}
   \begin{eqnarray}
	a_2^i &=& \frac{G}{r^2} \Bigg\{ n^i \Bigg[ -v^2
		5 \left( \frac{G M_{IBH}}{r} \right) \nonumber \\
	      &+& 4 \left( \frac{G M_{CBH}}{r} \right) \Bigg]
		+ 4 v^i n_j v^j \Bigg\}, \\
	\label{eq:PN1}
	a_4^i &=& \frac{G}{r^2} \Bigg\{ n^i \Bigg[
		\frac{G M_{IBH}}{r} \Bigg(-\frac{15}{4}v^2  
	        + \frac{39}{2} (n_j v^j)^2 \Bigg) \nonumber \\
	      &+& 2\frac{G M_{CBH}}{r}(n_j v^j)^2 \Bigg] -  v^i \frac{G}{r} n_j v^j \nonumber \\
	      &\times& \Bigg( \frac{63}{4} M_{IBH} 
	        + 2 M_{CBH} \Bigg) \Bigg\} \\
	      &+& \frac{G^3 M_{CBH}}{r^4} n^i \Bigg(-\frac{57}{4} M_{IBH}^2 
	        - 9 M_{CBH}^2 \nonumber \\
	      &-& \frac{69}{2} M_{IBH}M_{CBH} \Bigg), \nonumber \\
	\label{eq:PN2}
	a_5^i &=& \frac{4}{5} \frac{G^2 M_{IBH} M_{CBH}}{r^3} \nonumber \\
	      &\times& \Bigg\{v^i \Bigg[-v^2 + 2 \frac{G M_{IBH}}{r} \\
              &-& 8 \frac{G M_{CBH}}{r} \Bigg] + n^i n_j v^j \nonumber \\
              &\times& \Bigg[3v^2 - 6 \frac{G M_{IBH}}{r} + \frac{52}{3} \frac{G M_{CBH}}{r}
		\Bigg] \Bigg\}. \nonumber
	\label{eq:PN25}
   \end{eqnarray}
\end{mathletters}
where $\vec{n}$ is the unit vector pointing towards the CBH and the Einstein summation 
notation is being used.  Averaging changes in the semi-major axis and eccentricity 
due to the accelerate presented in equation (\ref{eq:PN25}) over orbital periods 
reproduces the familiar result presented by \citet{Peters64},
\begin{mathletters}
   \begin{eqnarray}
	\left\langle \frac{da}{dt} \right\rangle &=& \frac{64}{5} 
		\frac{G^2 M_{IBH} M_{CBH} \left(M_{IBH}+M_{CBH} \right)}
		{c^5a^3 \left(1-e^2 \right)^{7/2}} \nonumber \\
		&\times& \left(1 + \frac{73}{24}e^2 + \frac{37}{96}e^4 \right), \\
	\label{eq:dadt}
	\left\langle \frac{de}{dt} \right\rangle &=& \frac{304}{15}
		\frac{G^2 M_{IBH} M_{CBH} \left(M_{IBH}+M_{CBH} \right)}
		{c^5a^4 \left(1-e^2 \right)^{5/2}} \nonumber \\
		&\times& \left(1 + \frac{121}{304}e^2 \right),
	\label{eq:dedt}		
   \end{eqnarray}
\end{mathletters}
where $a$ is the semi-major axis and $e$ is the eccentricity \citep{Gultekin2006}.

\section{Orbital Characteristics}

Using the formalism defined above, we identify three broad categories of orbital history.
 The most
noticable differences between these regimes are the length of time needed for
the system to progress through the different stages, the effect of 
orbital resonances between the two
IBHs and the degree to which the relative inclination varies. 

\begin{figure}
 \centering
 \leavevmode
% \centering
%  \subfigure[Semi-major axis vs. time]{
   \includegraphics[scale=0.25, angle=0]{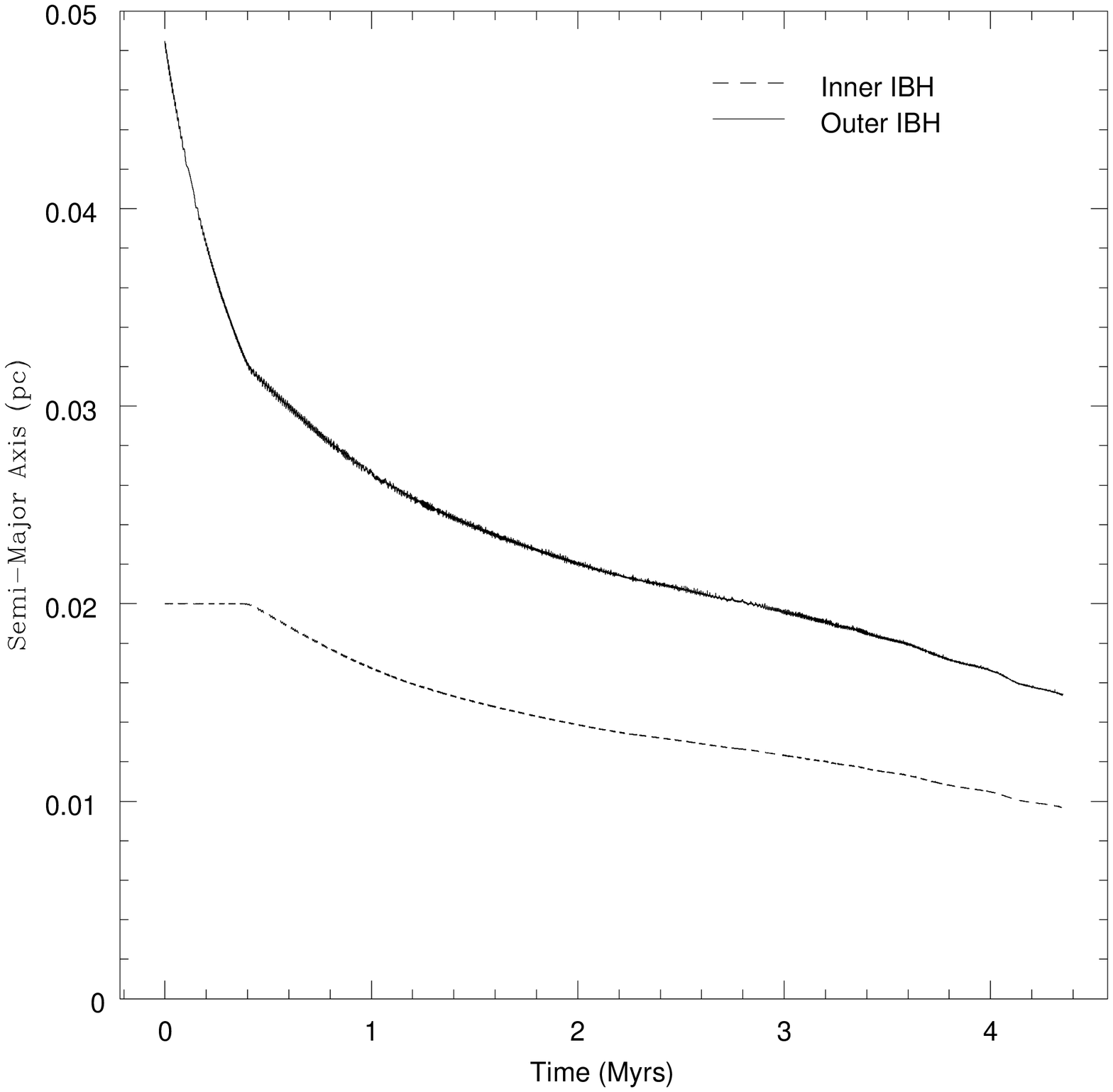}
%   \caption{Semi Major Axis vs. Time where an IBH merged with the CBH}
%  \hspace{.07in}
   \hfil
%  \subfigure[Semi-major axis vs. time]{
   \includegraphics[scale=0.25, angle=0]{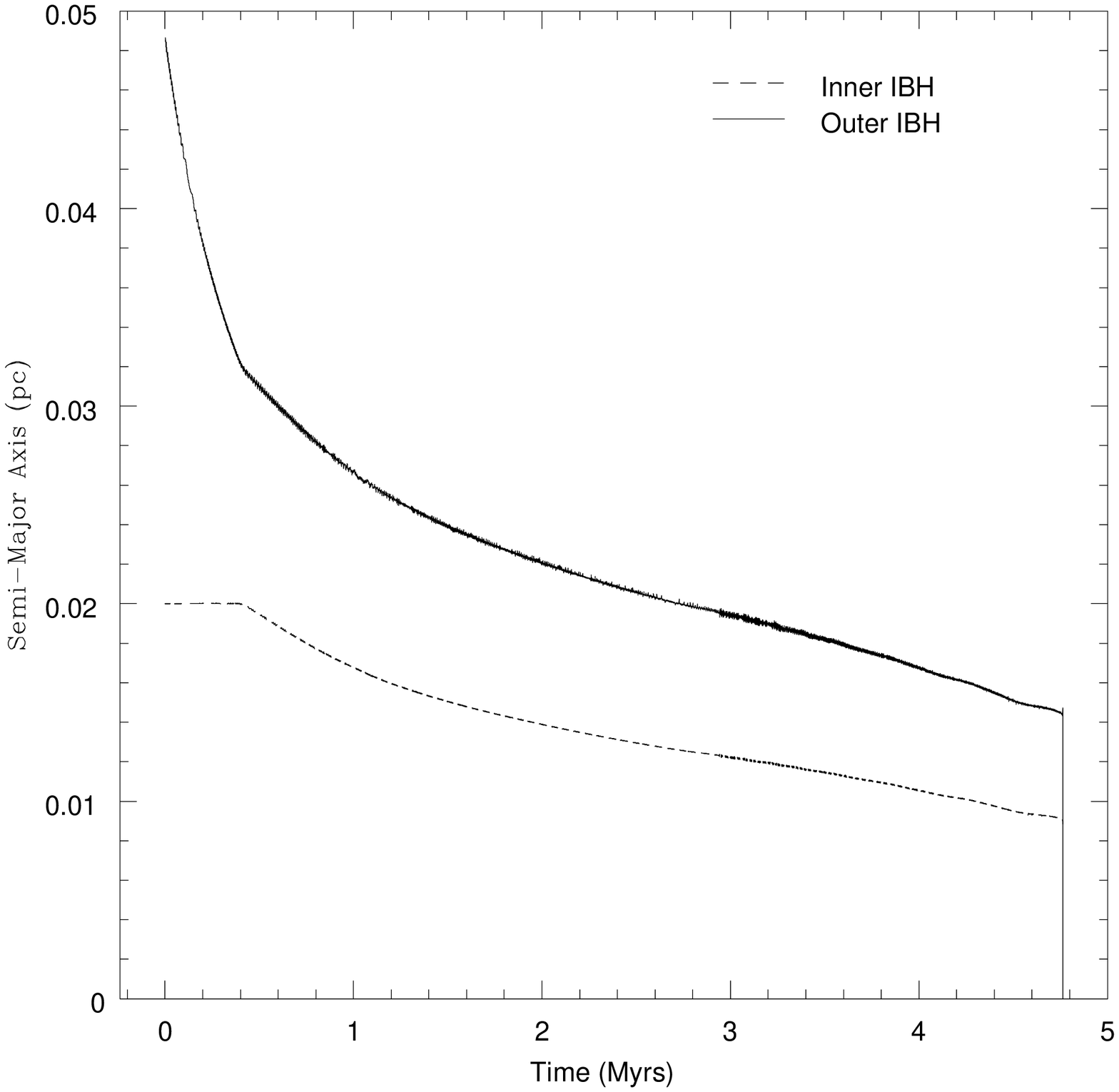}

%  \subfigure[Eccentricity vs. time]{
   \includegraphics[scale=0.25, angle=0]{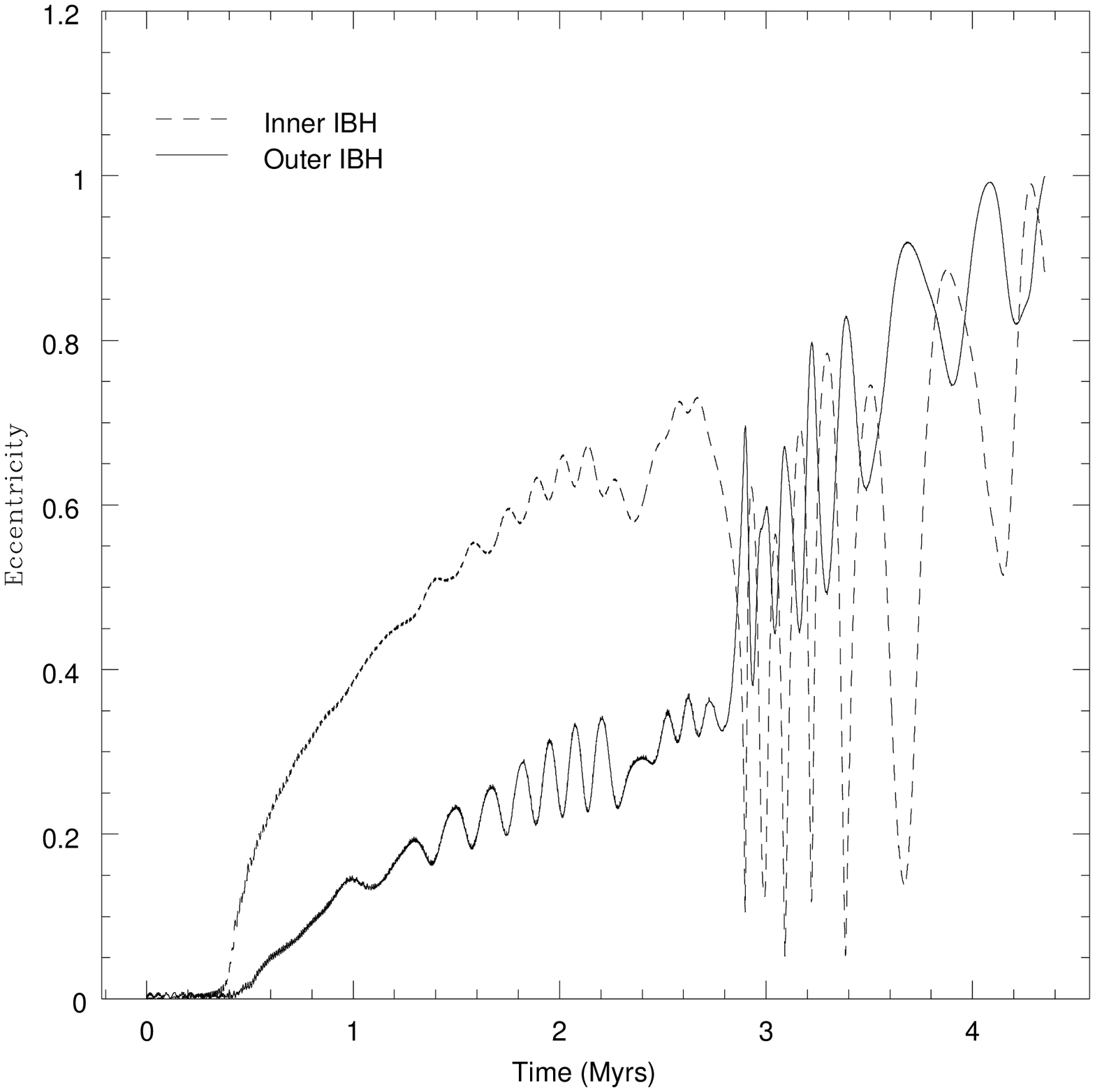}
  \hfil
%  \subfigure[Eccentricity vs. time]{
   \includegraphics[scale=0.25, angle=0]{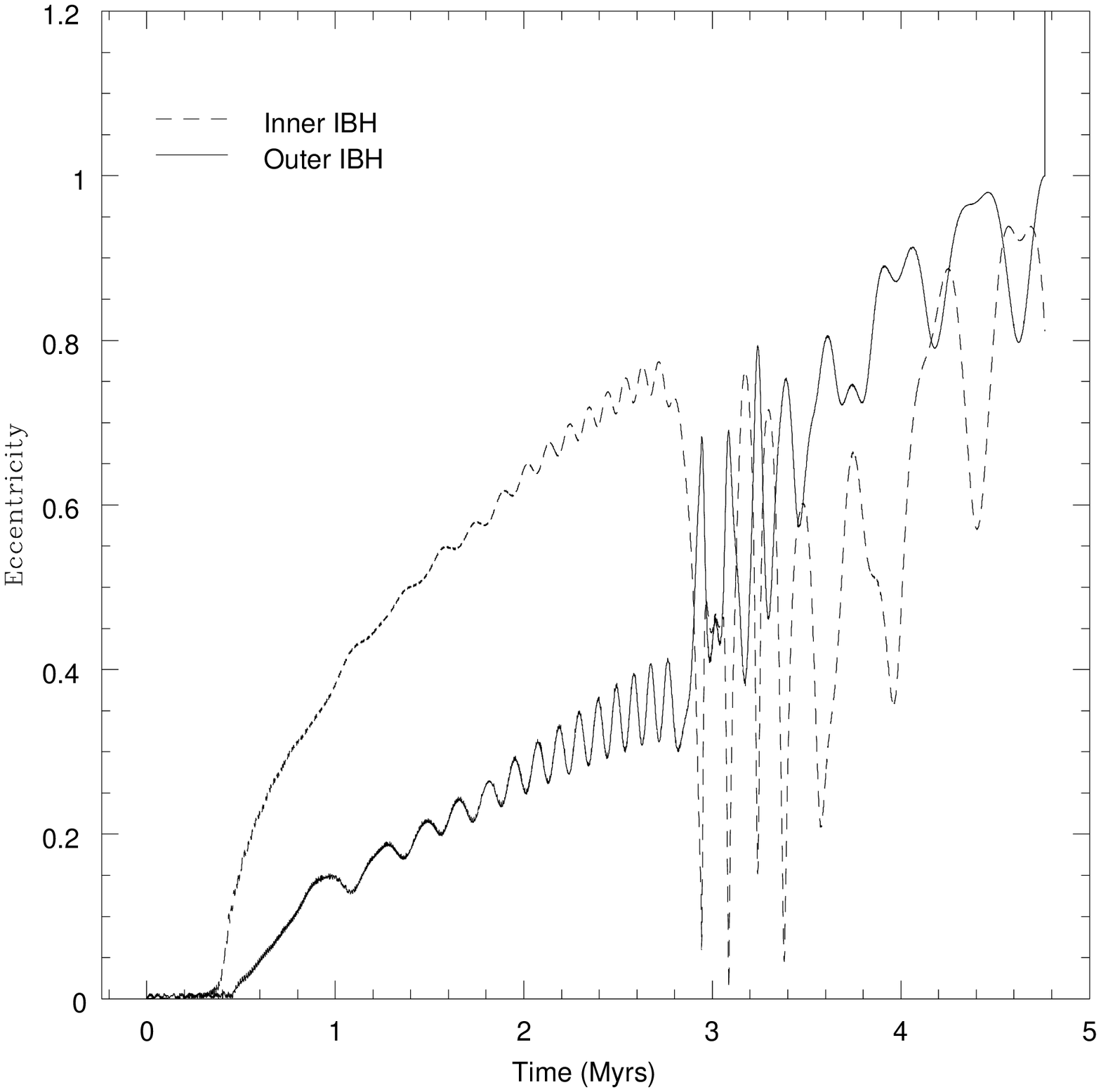}

%  \subfigure[Angle of inclination vs. time]{
   \includegraphics[scale=0.25, angle=0]{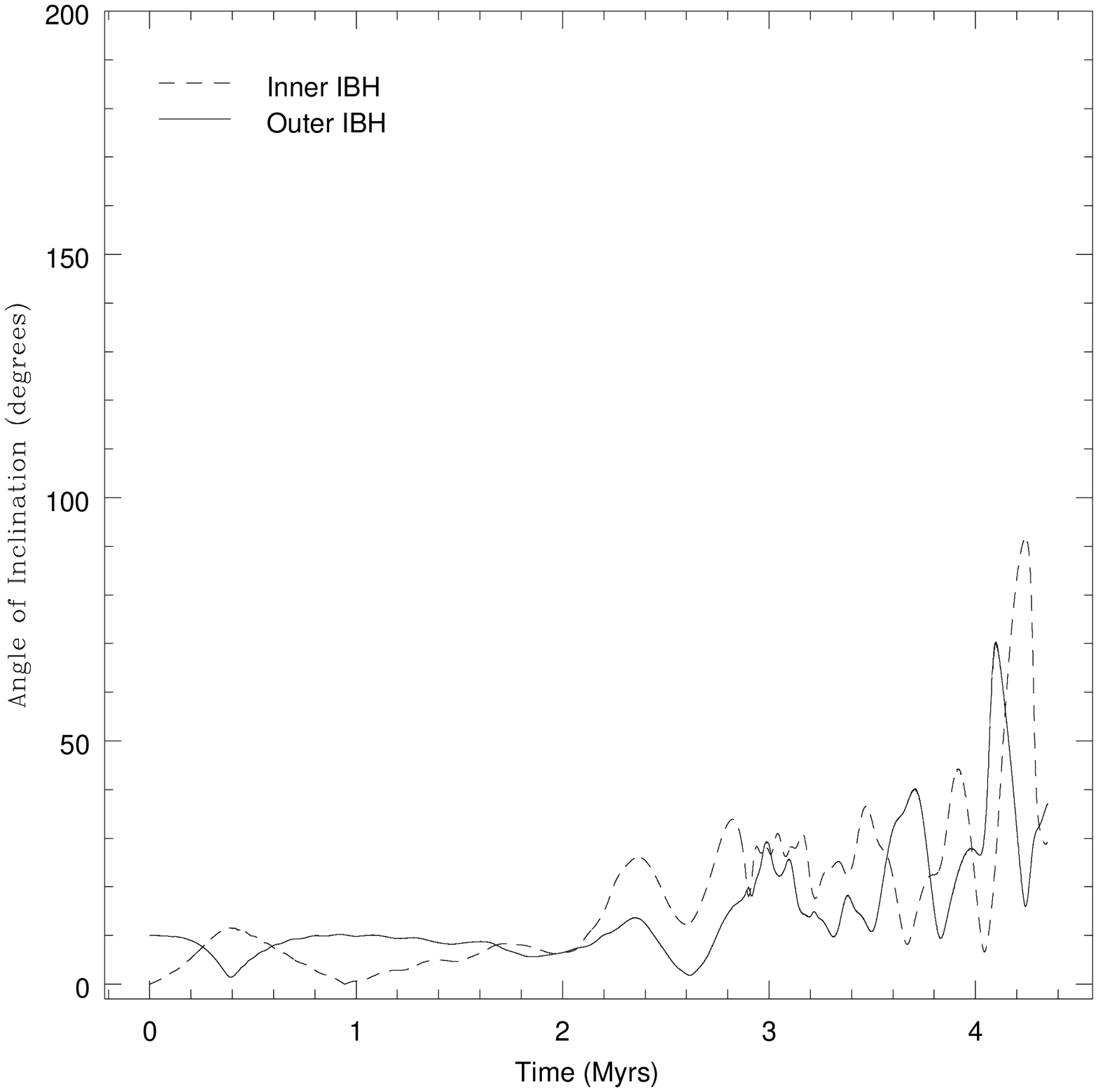}
  \hfil
%  \subfigure[Angle of inclination vs. time]{
   \includegraphics[scale=0.25, angle=0]{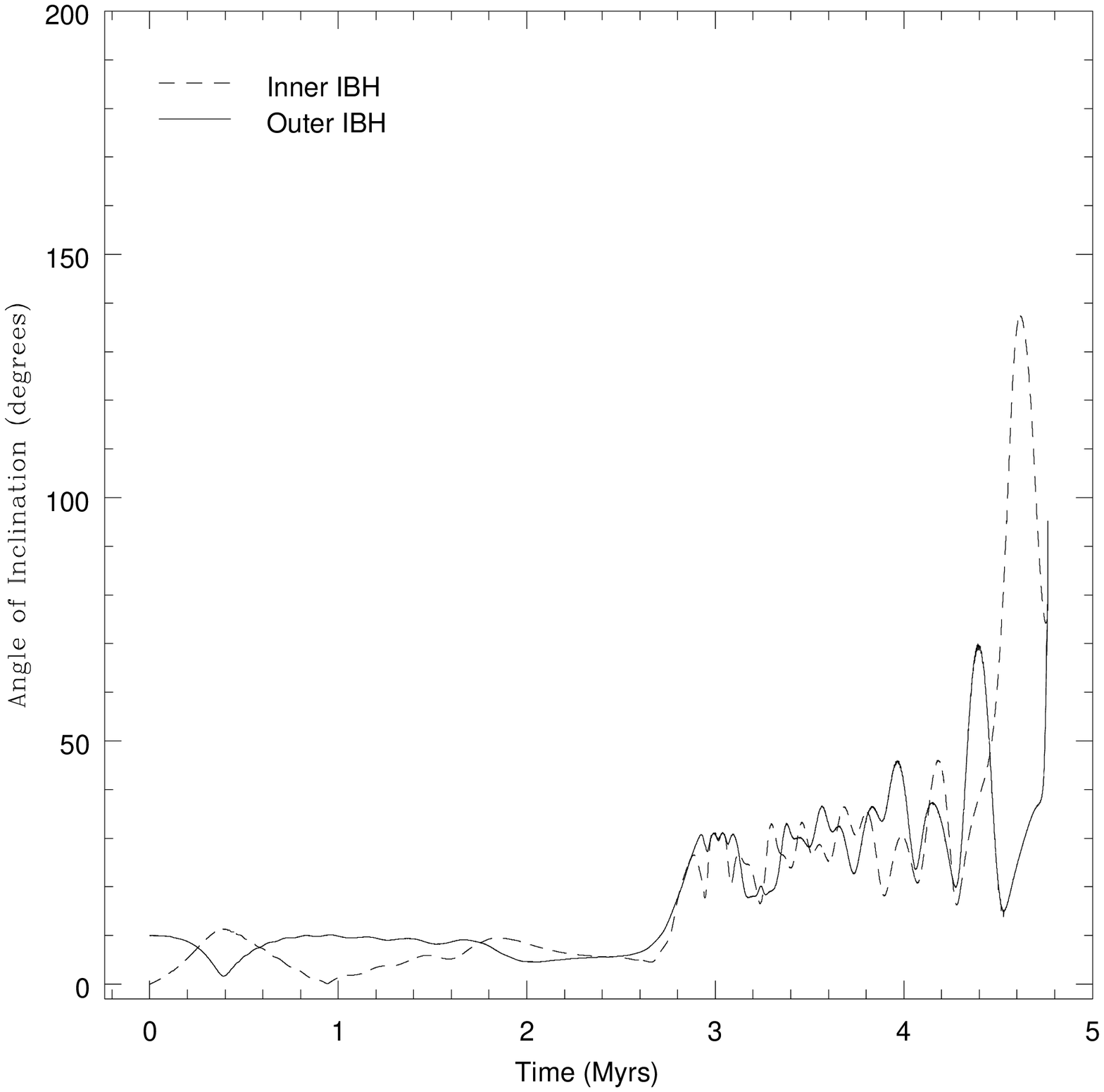}

 \caption{Orbital parameters, semi-major axis (upper left merged, 
          upper right ejected), eccentricity (middle left merged, 
          middle right ejected) and angle of inclination (lower left 
          merged, lower right ejected) as a function of time for 
          $\Delta i_0 = 10^{\o}$.
          The result of these simulations were that the outer IBH merged
          with the CBH for plots on the left and the outer IBH was
          ejected beyond $4 \pc$ for plots on the right.}
 \label{fig:orbparam10}
\end{figure}

The first category consists of prograde orbits, with a small
initial angle of inclination, $0\dgrs < \Delta i_0 < 40\dgrs$.
Figure (\ref{fig:orbparam10}) is a plot of the orbital parameters
for both of the IBHs as a function of time for $\Delta i_0 = 10^{\o}$.
This plot shows that the
two IBHs first begin to interact with each other at about 0.5 $\pm 0.05$ Myrs,
which is roughly the same for all prograde orbits. The two IBH soon become
locked in a 2:1 orbital mean-motion resonance, which leads to the orbit
of the inner IBH beginning to decay as well. The overall rate of decay slows,
since the strong mutual interaction means that the
 energy extracted by dynamical friction on the outer IBH
is removed from both IBH orbits in concert. In fact, as the evolution proceeds,
and the
eccentricities of both orbits grow, the apastron of the inner IBH can reach out beyond
the evacuated region and thus 
both IBH  experience
some measure of dynamical friction. Continued evolution leads to eccentricity
growth and eventually, after $\sim 4.5$ Myr, the eccentricities and inclinations
experience large-amplitude variations, leading to instability. The time at
which this epoch of strong interactions occurs depends on the initial relative
inclination, occurring earlier for larger values of $ \Delta i_0 $. The two
IBH continue to interact until either one is ejected or passes sufficiently
close to the CBH that it loses orbital energy to gravitational radiation,
resulting in an eventual merger with the CBH.

\begin{figure}
 \centering
 \leavevmode
% \plotone{avst40.eps}
% \plottwo{evst40.eps}{ivst140.eps}

%  \subfigure[Semi-major axis vs. time]{
   \includegraphics[scale=0.25, angle=0]{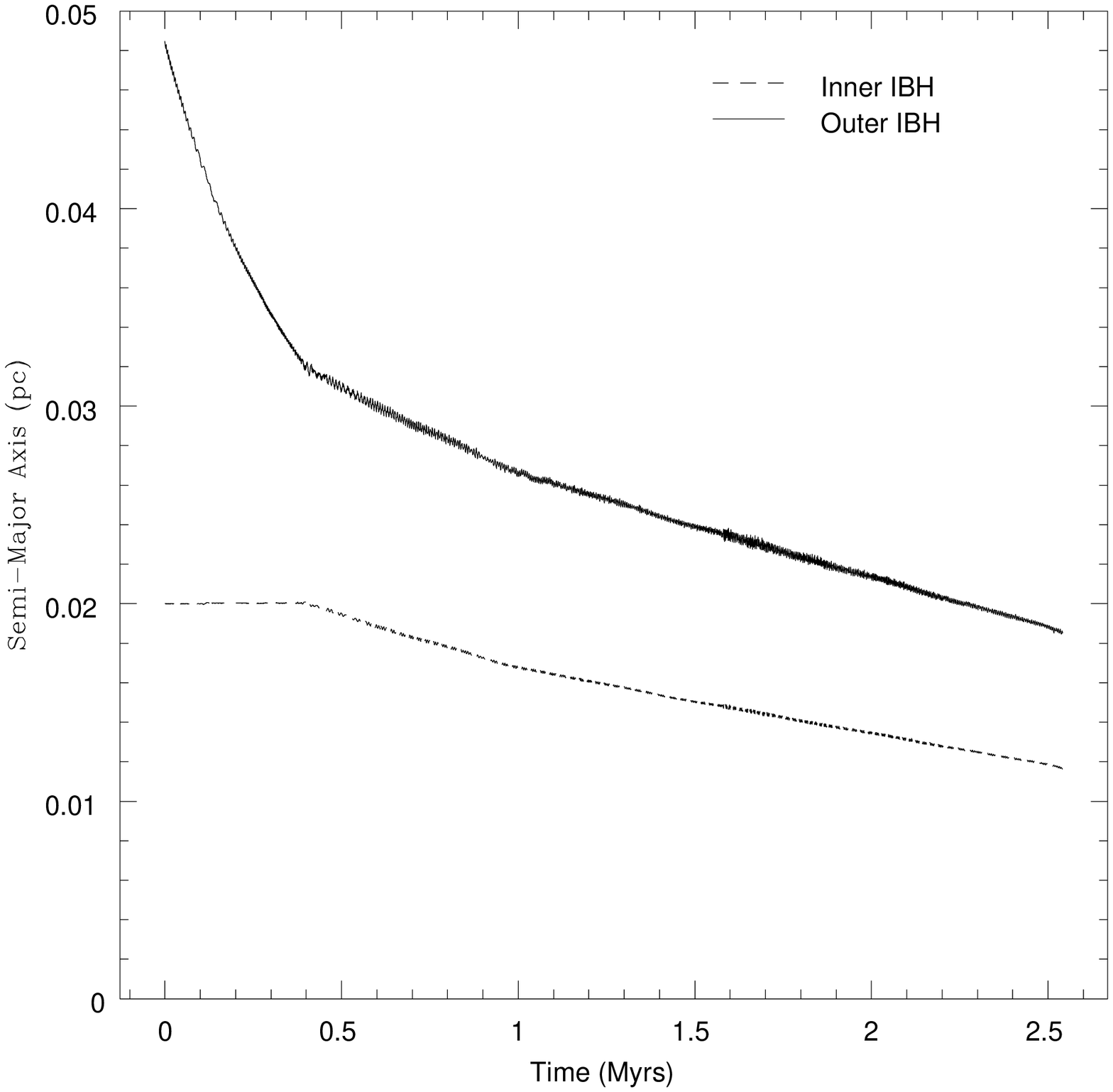}
  \hfil
%  \subfigure[Semi-major axis vs. time]{
   \includegraphics[scale=0.25, angle=0]{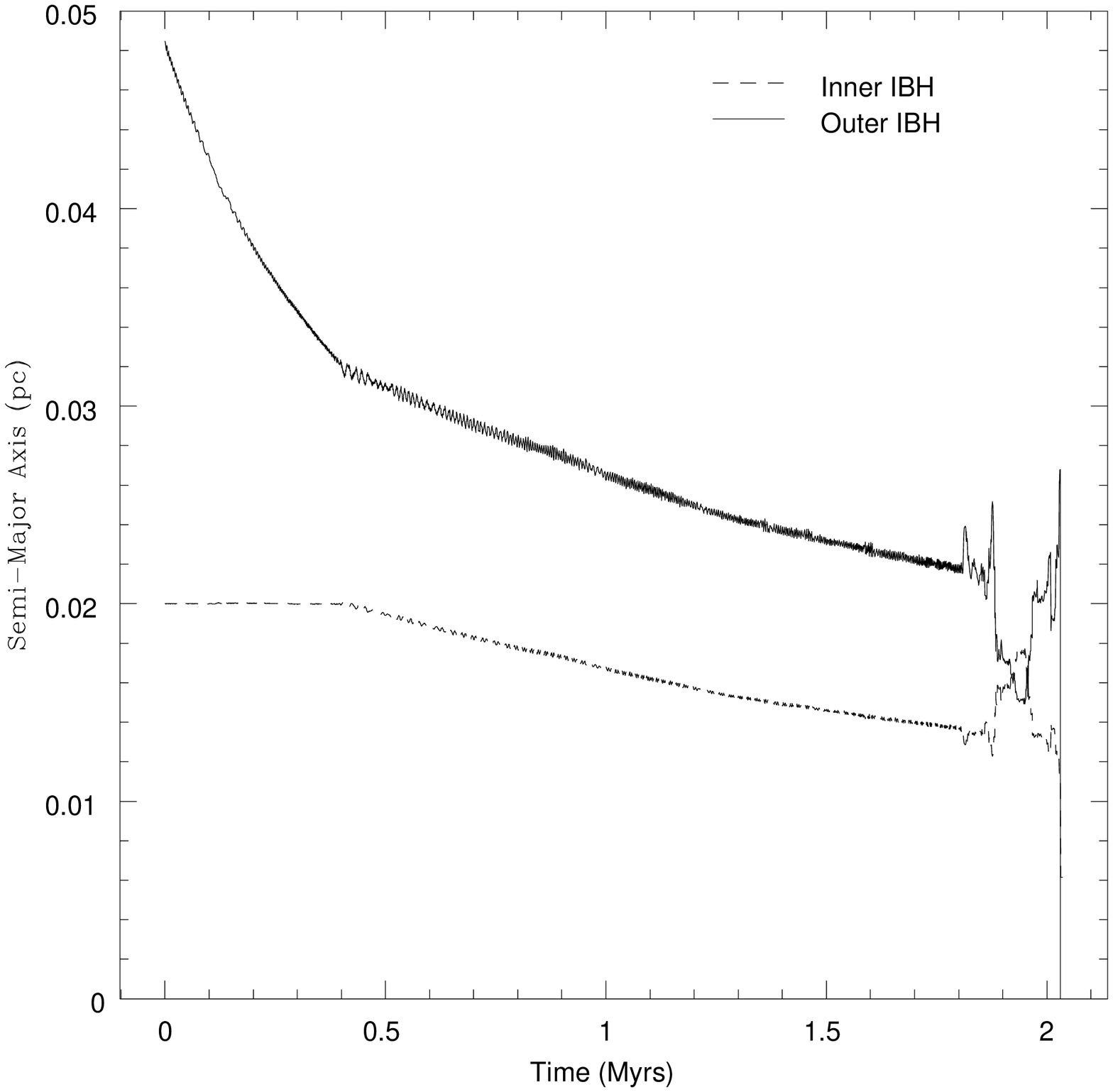}

%  \subfigure[Eccentricity vs. time]{
   \includegraphics[scale=0.25, angle=0]{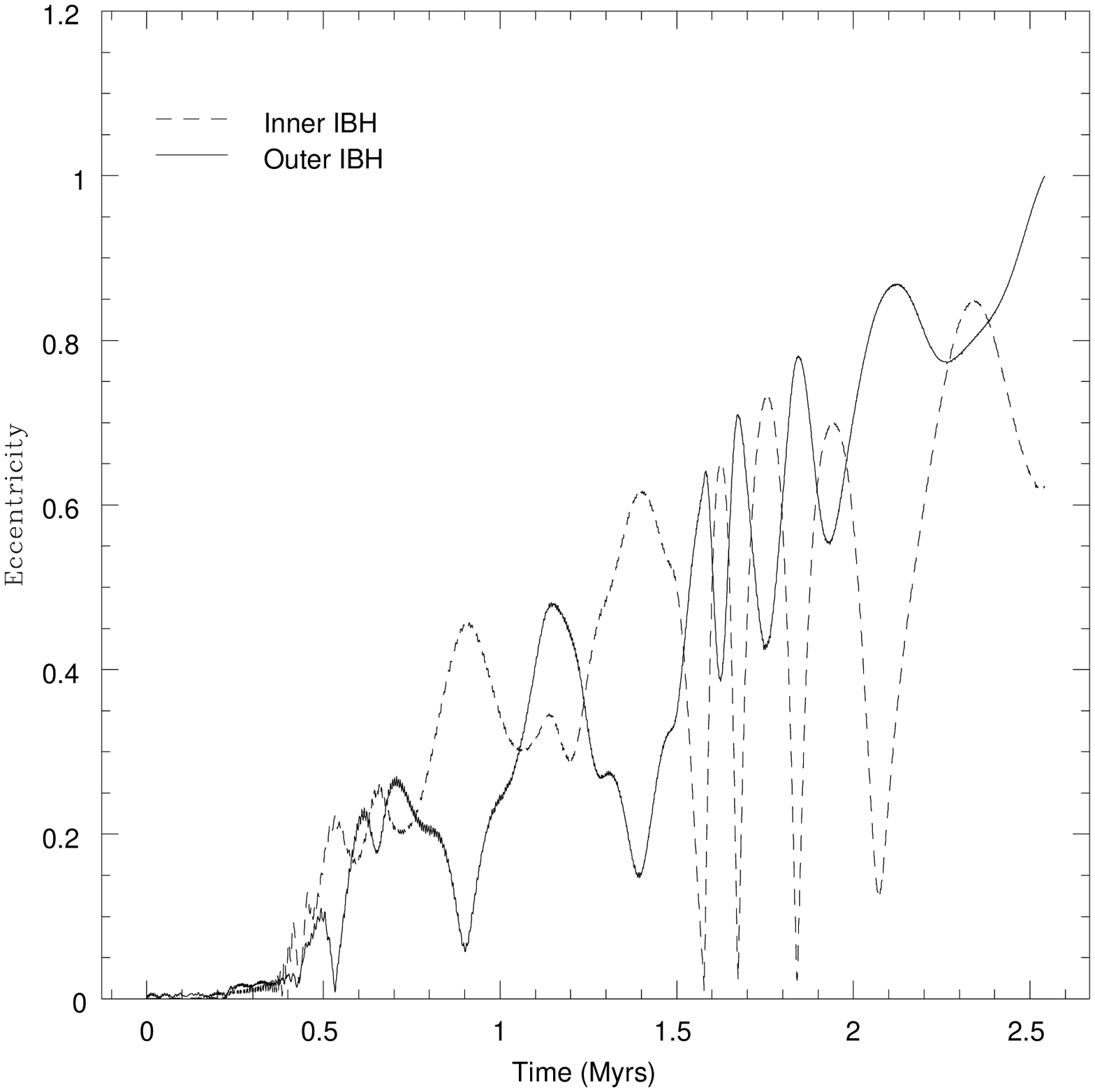}
  \hfil
%  \subfigure[Eccentricity vs. time]{
   \includegraphics[scale=0.25, angle=0]{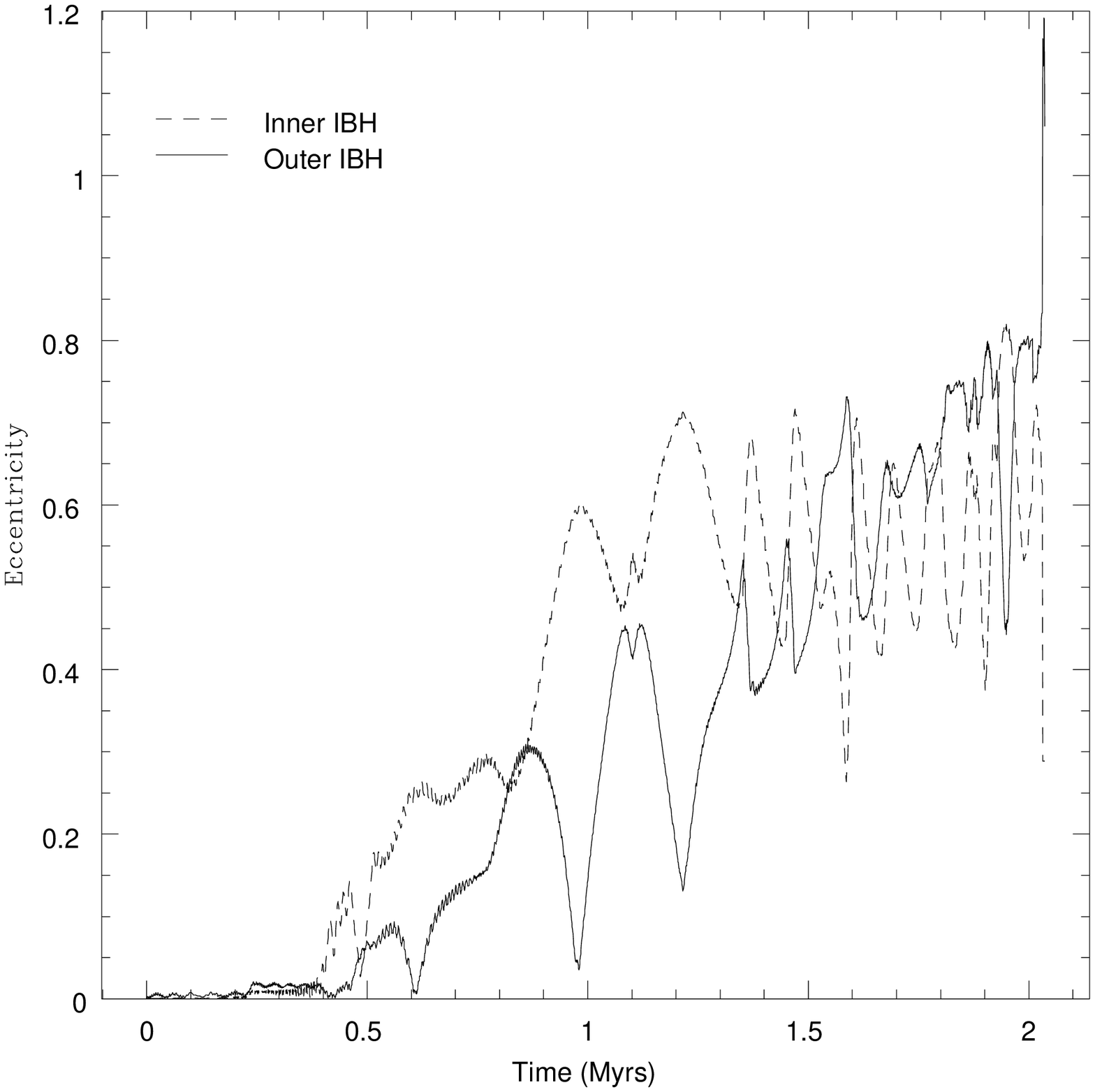}

%  \subfigure[Angle of inclination vs. time]{
   \includegraphics[scale=0.25, angle=0]{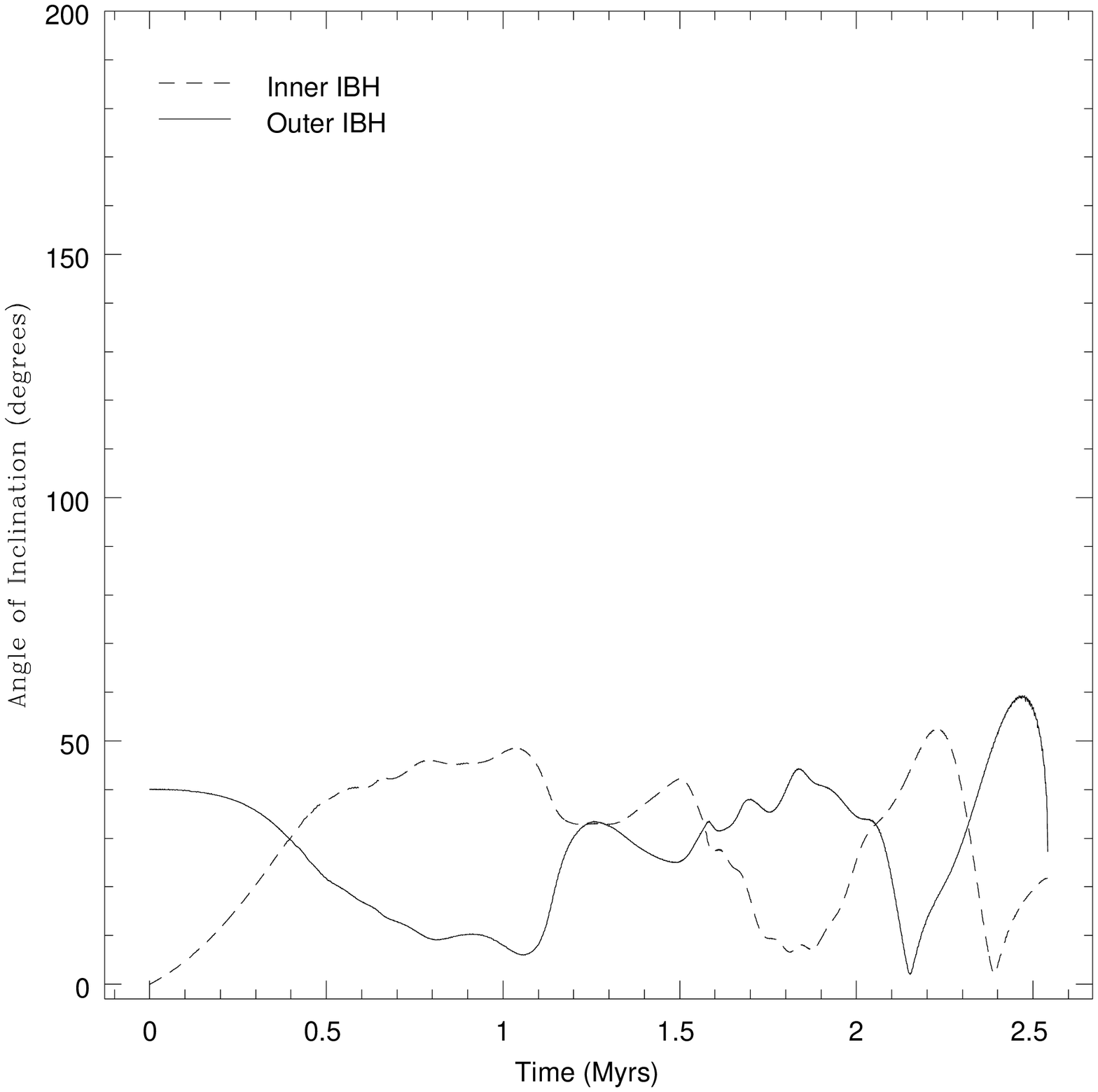}
  \hfil
%  \subfigure[Angle of inclination vs. time]{
   \includegraphics[scale=0.25, angle=0]{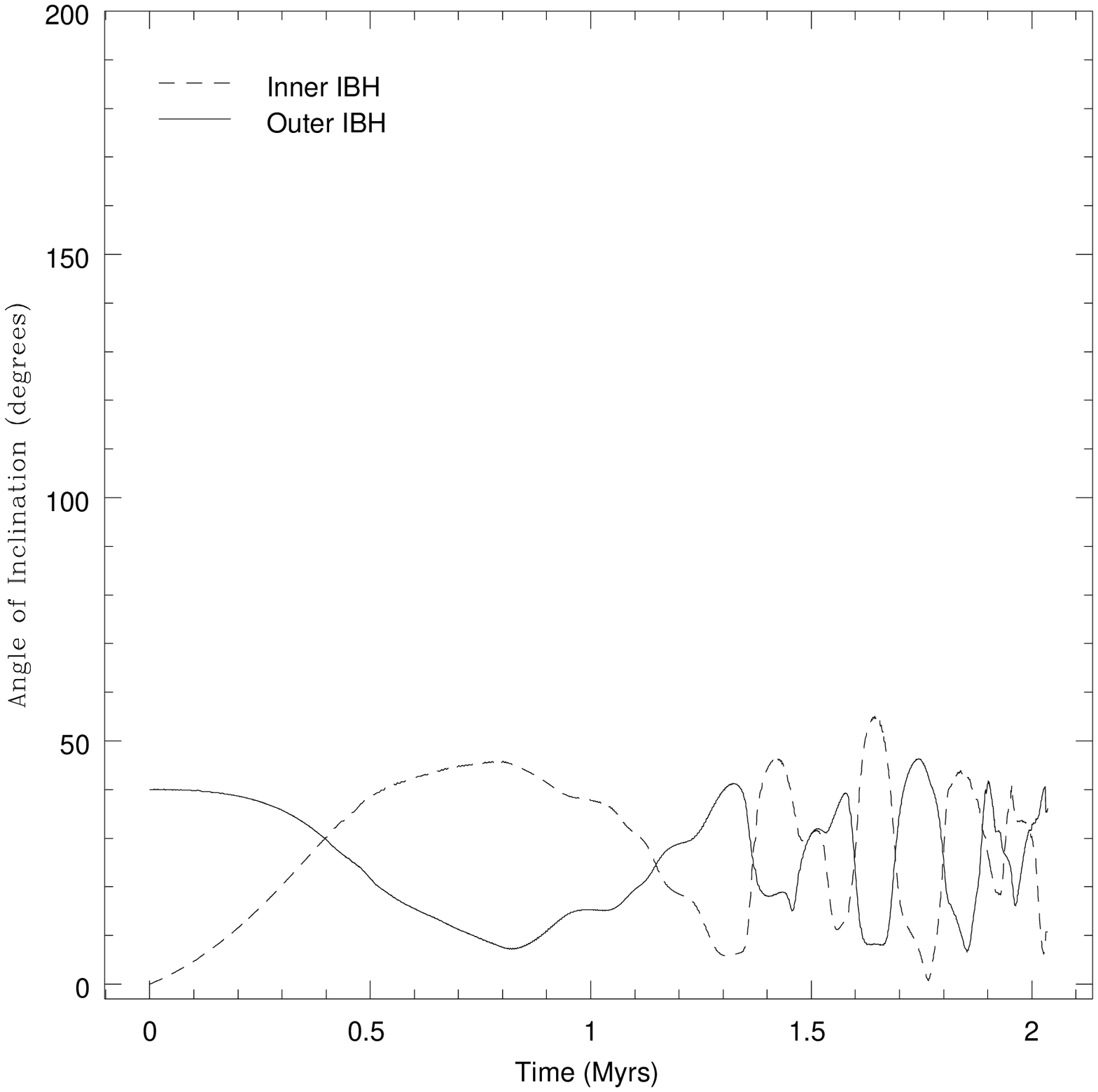}

 \caption{Orbital parameters, semi-major axis (upper left merged,
          upper right ejected), eccentricity (middle left merged,
          middle right ejected) and angle of inclination (lower left
          merged, lower right ejected) as a function of time for
          $\Delta i_0 = 40^{\o}$.
          The result of these simulations were that the outer IBH merged
          with the CBH for plots on the left and the outer IBH was
          ejected beyond $4 \pc$ for plots on the right.}
 \label{fig:orbparam40}
\end{figure}

As $\Delta i_0$ increases, the time that elapses before strong interactions
occur decreases. 
This rarely occurs before 2 Myrs for $\Delta i_0 < 40 \dgrs$.
However, for $\Delta i_0 \ge 40 \dgrs$ (except $\Delta i_0 = 180 \dgrs$)
strong  perturbations almost immediately start affecting the orbit,
presumably as a result of the Kozai effect\citep{Kozai62} (also
seen in N-body simulations, e.g. \citet{Aar07}). We classify the orbits 
with  $40\dgrs \le \Delta i_0 \le 90 \dgrs$ as our second class of
orbital solutions.
Figure (\ref{fig:orbparam40}) contains plots of the evolution of orbital
parameters for $\Delta i_0 = 40 \dgrs$ and shows the rapid onset
of eccentricity and inclination growth.
Like the prograde orbits with $\Delta i_0 < 40 \dgrs$, a 2:1 orbital resonance
is established at about 0.5 Myrs, but shortly afterward, at about 1.5 Myrs,
strong oscillations
begin to dominate the resonant behavior of the eccentricity and angle of
inclination.

\begin{figure}
 \centering
 \leavevmode
% \plotone{avst140.eps}
% \plottwo{evst140.eps}{ivst140.eps}

%  \subfigure[Semi-major axis vs. time]{
   \includegraphics[scale=0.25, angle=0]{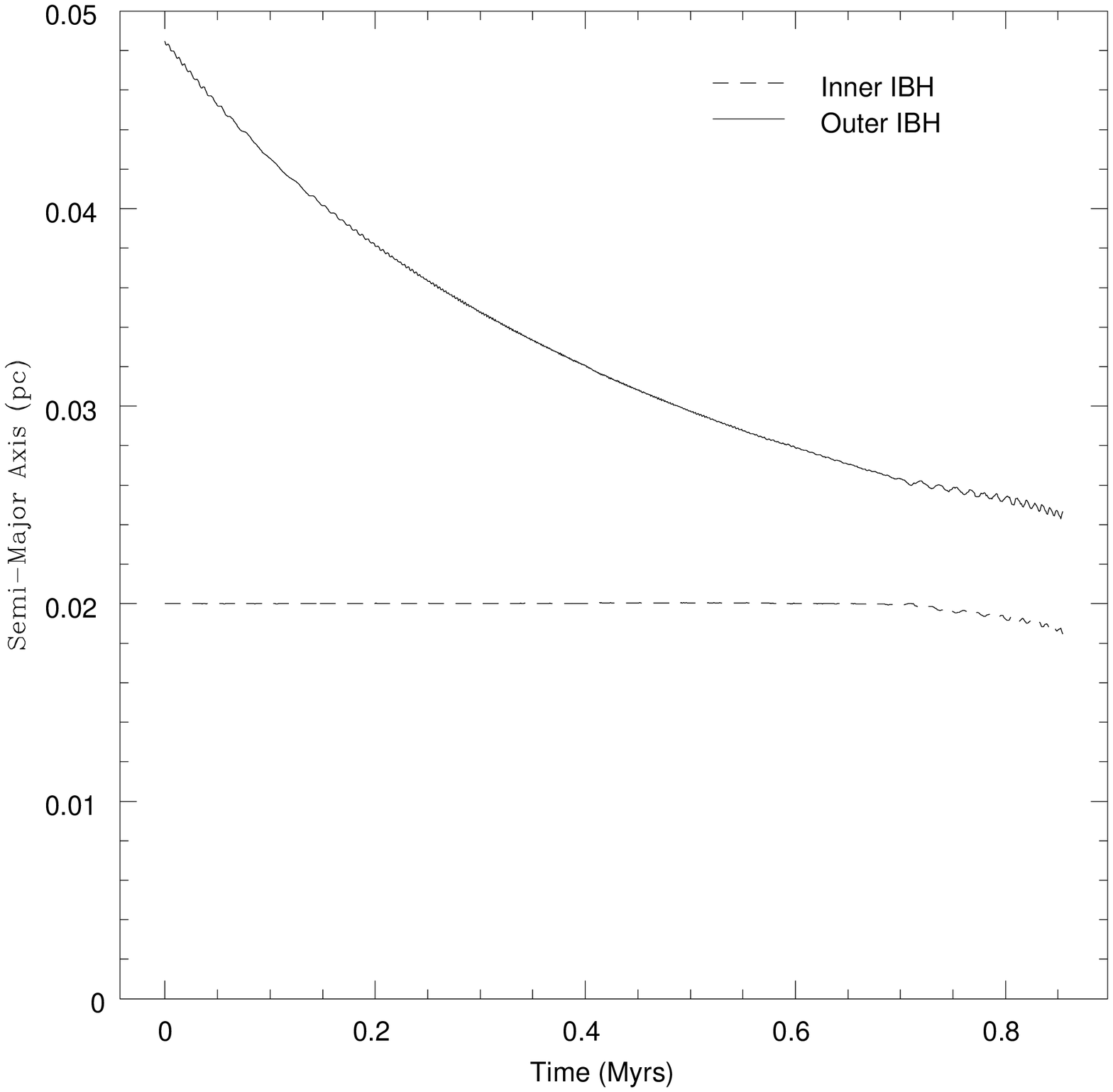}
  \hfil
%  \subfigure[Semi-major axis vs. time]{
   \includegraphics[scale=0.25, angle=0]{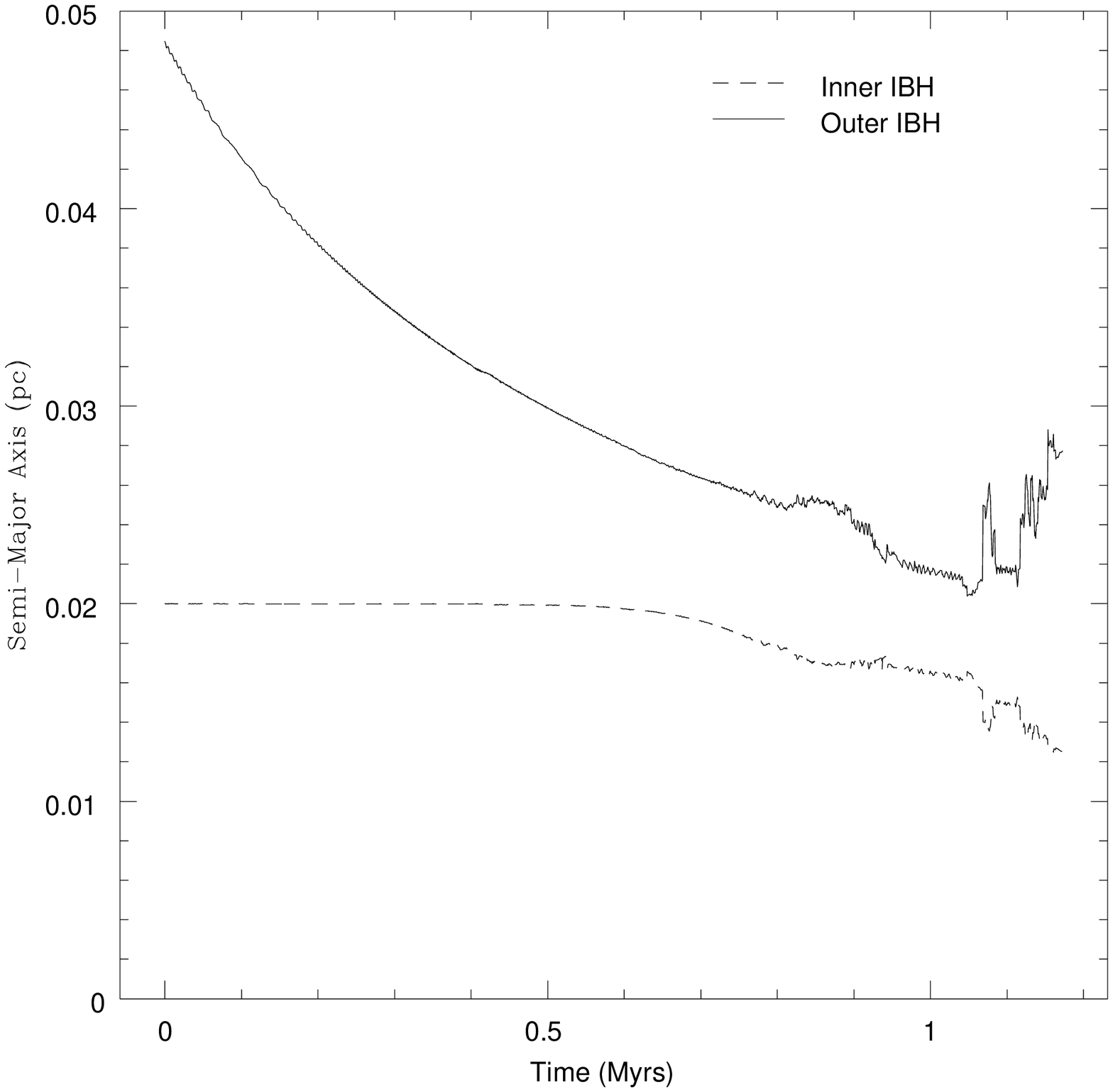}

%  \subfigure[Eccentricity vs. time]{
   \includegraphics[scale=0.25, angle=0]{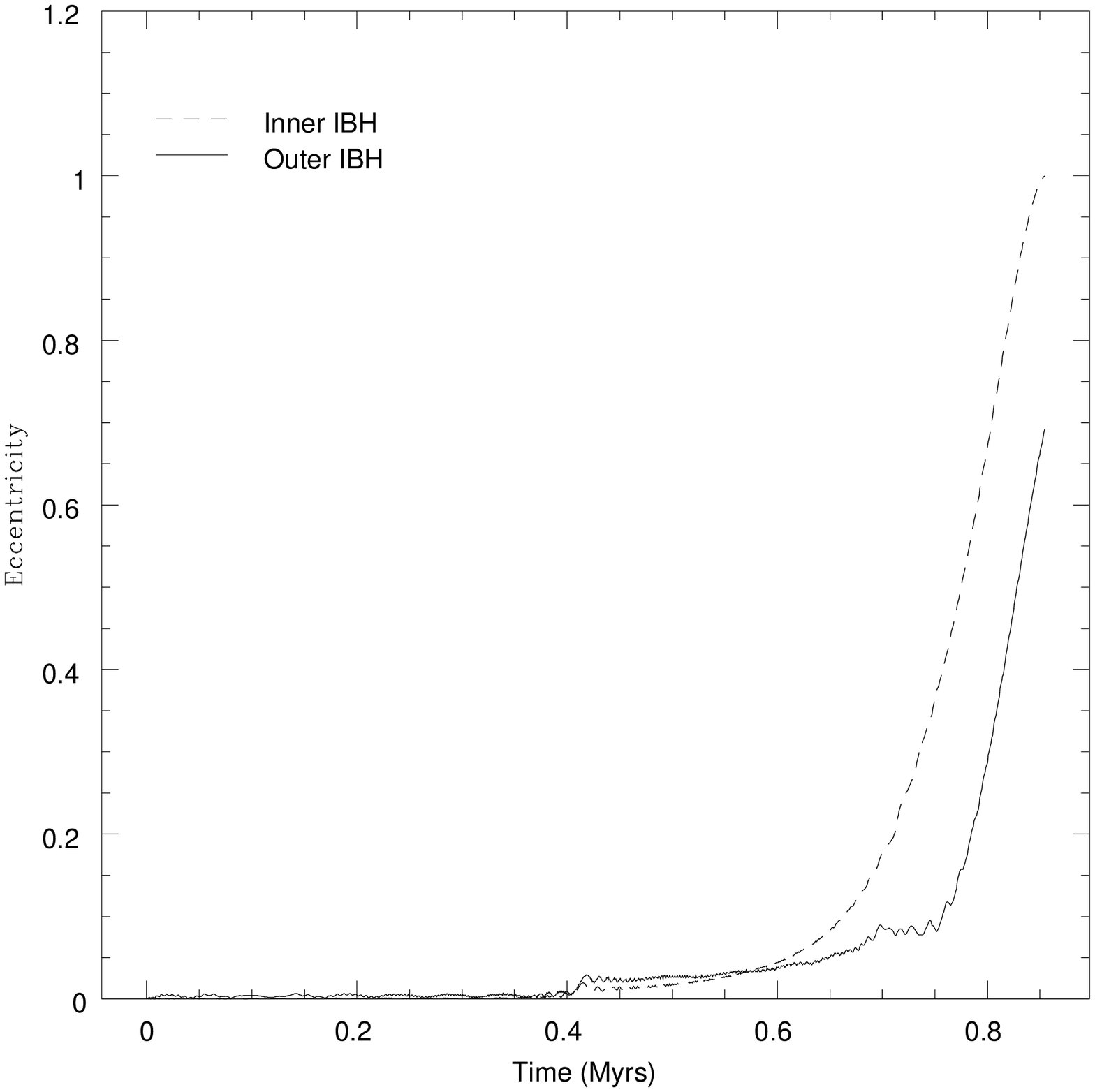}
  \hfil
%  \subfigure[Eccentricity vs. time]{
   \includegraphics[scale=0.25, angle=0]{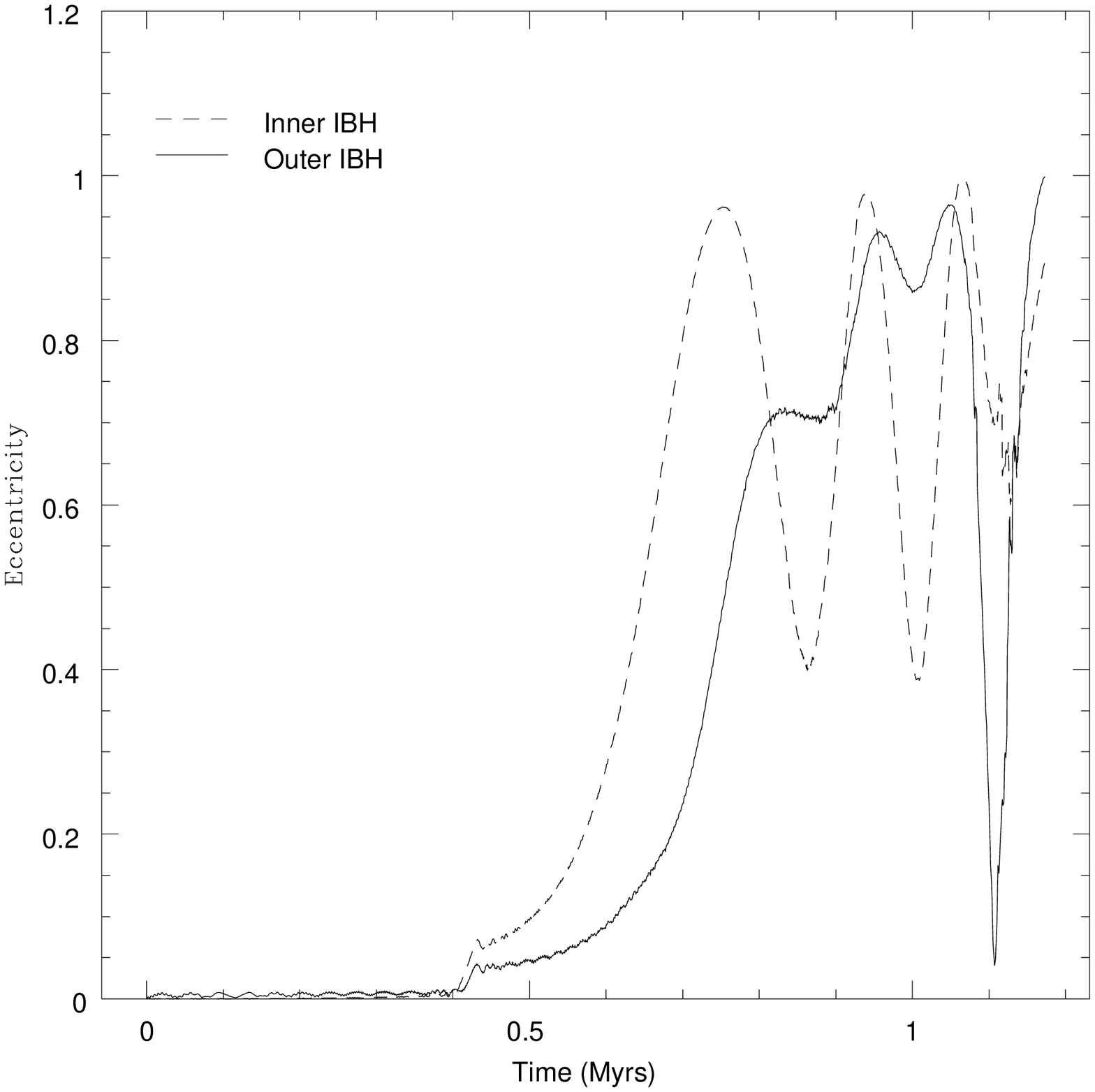}

%  \subfigure[Angle of inclination vs. time]{
   \includegraphics[scale=0.25, angle=0]{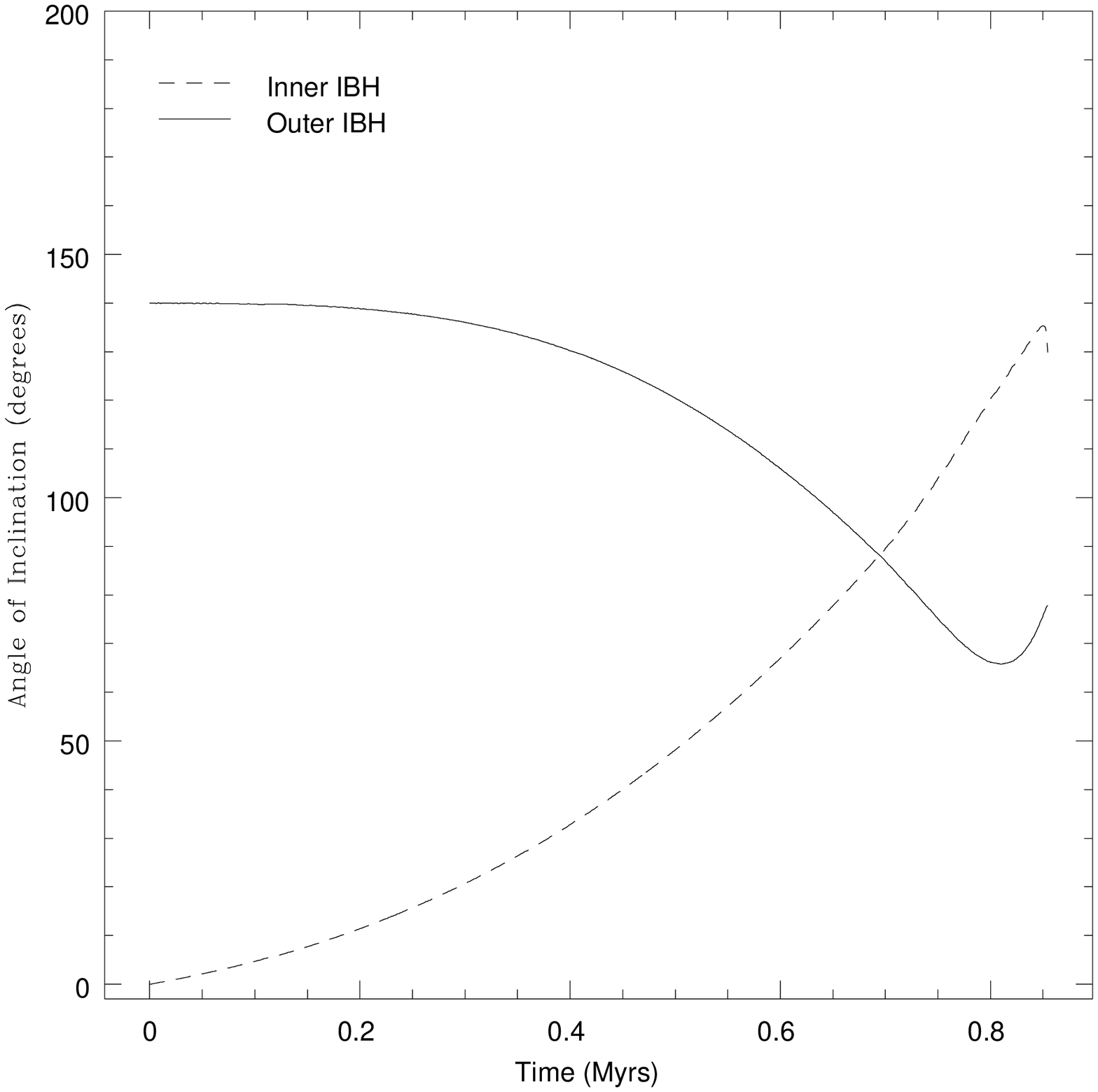}
  \hfil
%  \subfigure[Angle of inclination vs. time]{
   \includegraphics[scale=0.25, angle=0]{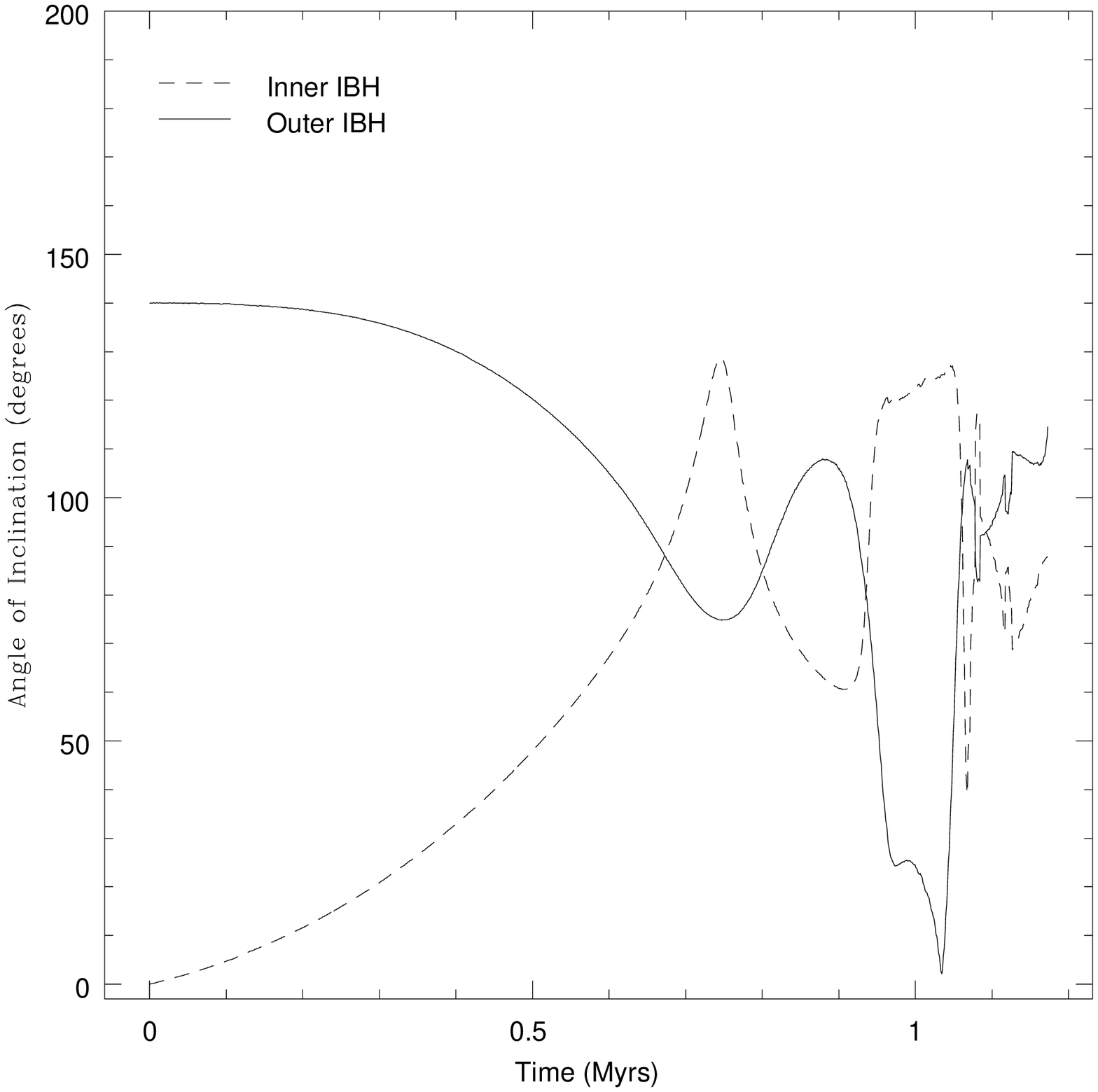}

 \caption{Orbital parameters, semi-major axis (upper left merged,
          upper right ejected), eccentricity (middle left merged,
          middle right ejected) and angle of inclination (lower left
          merged, lower right ejected) as a function of time for
          $\Delta i_0 = 140^{\o}$.
          The result of these simulations were that the inner IBH merged
          with the CBH for plots on the left and the outer IBH was
          ejected beyond $4 \pc$ for plots on the right.}
 \label{fig:orbparam140}
\end{figure}

In the third regime, $\Delta i_0 > 90 \dgrs$, strong oscillations in
inclination begin almost immediately
(figure \ref{fig:orbparam140}).  
Additionally, since the two orbits are retrograde
 with respect to each other,
the interaction times are
shorter and less frequent, therefore an orbital resonance is more
difficult to establish and the exchange of energy 
through ``close encounters'' is more important. However, there are
clearly still strong oscillations in eccentricity and inclination
most likely driven by the Kozai effect
 (figure \ref{fig:orbparam140}).  Although the result of the simulation
is essentially the same as the other initial inclination regimes,
how the result is achieved is obviously slightly different.  

The three groups of initial angles of inclination differ in the 
overall time periods associated with the evolution of an IBH's orbit,
but they all result in orbits that transfer angular momenta between the two
IBHs as seen through large oscillations in inclination angle and eccentricity
of the IBHs' orbits.  Ultimately the transfer of orbital energy and 
angular momentum between the two IBH results in either an IBH-CBH merger 
or the ejection of one of the IBH.
Mergers generally occur when the resonant interactions drive the
 eccentricity of one of the IBH 
to the point where it starts to interact strongly with the CBH near periastron,
radiating gravitational waves and 
resulting eventually in a merger. Ejections occur when the resonant avoidance
fails and the two IBH undergo close encounters. Eventually the 
IBH undergo a close enough passage that one of them is scattered
onto a hyperbolic orbit. The end result in both cases is to leave the 
remaining IBH on an eccentric orbit.

\subsection{IBH-CBH Mergers, Ejections and 
            the Orbital Characteristics of the
            Remaining IBH}
\label{sec:MergeEject}

Although the majority of IBH-CBH mergers resulted in a gradual transfer 
and loss of energy such that the remaining IBH was left still orbiting the CBH, 
there appeared to be a small subset of IBH-CBH mergers, $\lesssim 1 \%$,
in which the other IBH was unbound.  Unlike the majority
of the IBH-CBH mergers (described below), these were the result of an
IBH-IBH scattering event that transfered enough energy from one of the IBHs to cause
the other to be ejected.  The IBH that remained bound was sufficiently eccentric that
it radiated the 
remainder of its orbital energy as gravitational radiation and merged with the CBH.

\begin{figure}
 \plotone{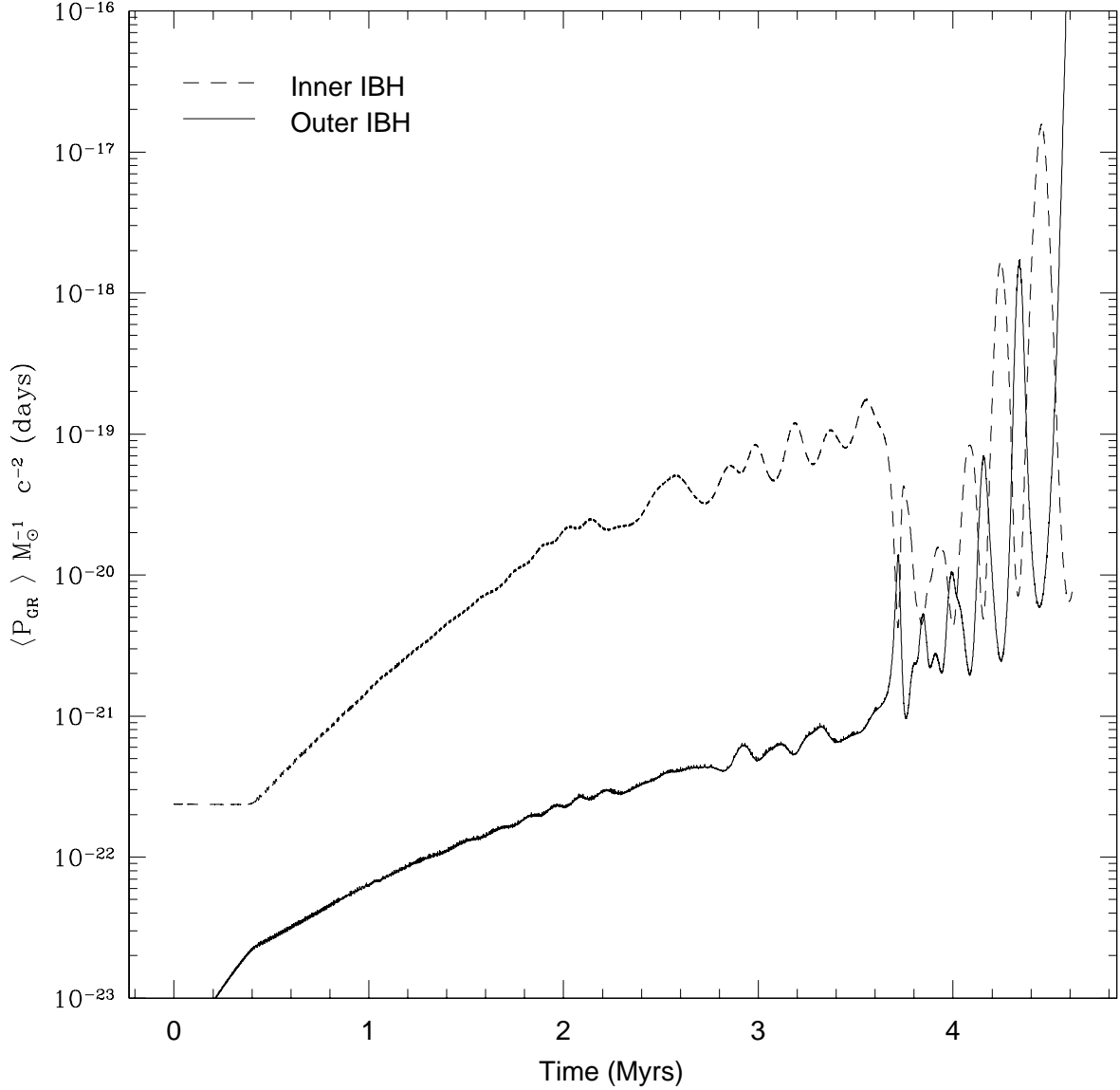}
 \caption{Normalised power radiated by gravitational radiation, 
          $\langle P_{GR} \rangle / {\rm\,M}_{\odot} c^2$, 
          as a function of time in Myrs for
          the two IBH.  Although this particular plot is associated with
	  an IBH-CBH merger with $\Delta i_0 = 10^\o$
	  (orbital parameters shown in figure \ref{fig:orbparam10})
          , it is a good 
	  representation of the typical power radiated as the IBH-IBH-CBH
	  three body system evolves, regardless of the final outcome.  The oscillatory behavour of 
	  $\langle P_{GR} \rangle / {\rm M}_{\odot c^2}$ corresponds with 
	  the transfer of angular momentum as well as orbital energy 
	  between the two IBHs.  As their eccentricities grow and 
	  periapsis decreases, more and more energy is radiated away until
	  there is one last, large release of energy before the IBH-CBH
	  merger occurs.}
 \label{fig:dedt}
\end{figure}

The eventual outcome of the dynamical interactions is either that one of the IBH 
is ejected from the immediate vicinity of the CBH, or that one of the IBH merges with the 
CBH.
Typically an IBH-CBH merger is the result of an IBH on a highly eccentric orbit
passing too closely to the CBH, whereby orbital energy from the 
IBH is carried away via gravitational radiation and the IBH spirals down.  
 Prior to the merger, orbital energy
and angular momentum are transferred between the two IBHs as they
orbit the CBH (and extracted by dynamical friction from the outermost IBH).
 This transfer of energy and angular momentum cause
the eccentricities of the orbits to oscillate.  Formally then, each IBH 
sends a burst of gravitational radiation as they pass through periastron.
As the periastron distance starts to approach the CBH, this energy 
loss becomes significant.
Figure (\ref{fig:dedt}) is a plot of the average power radiated through
gravitational waves and shows this oscillatory behavior.
Furthermore, this plot shows that the power radiated increases as the
orbits of the IBHs decay.  Lastly, to ensure that a IBH-CBH merger wasn't
a numerical artifact of the integration, the data in figure (\ref{fig:dedt})
was numerically integrated.  The resulting energy lost along with the 
energy lost to dynamical friction was within $0.1 \pm 0.05 \%$ of the
orbital energy of the IBH that merged. 

\begin{figure}
 \plotone{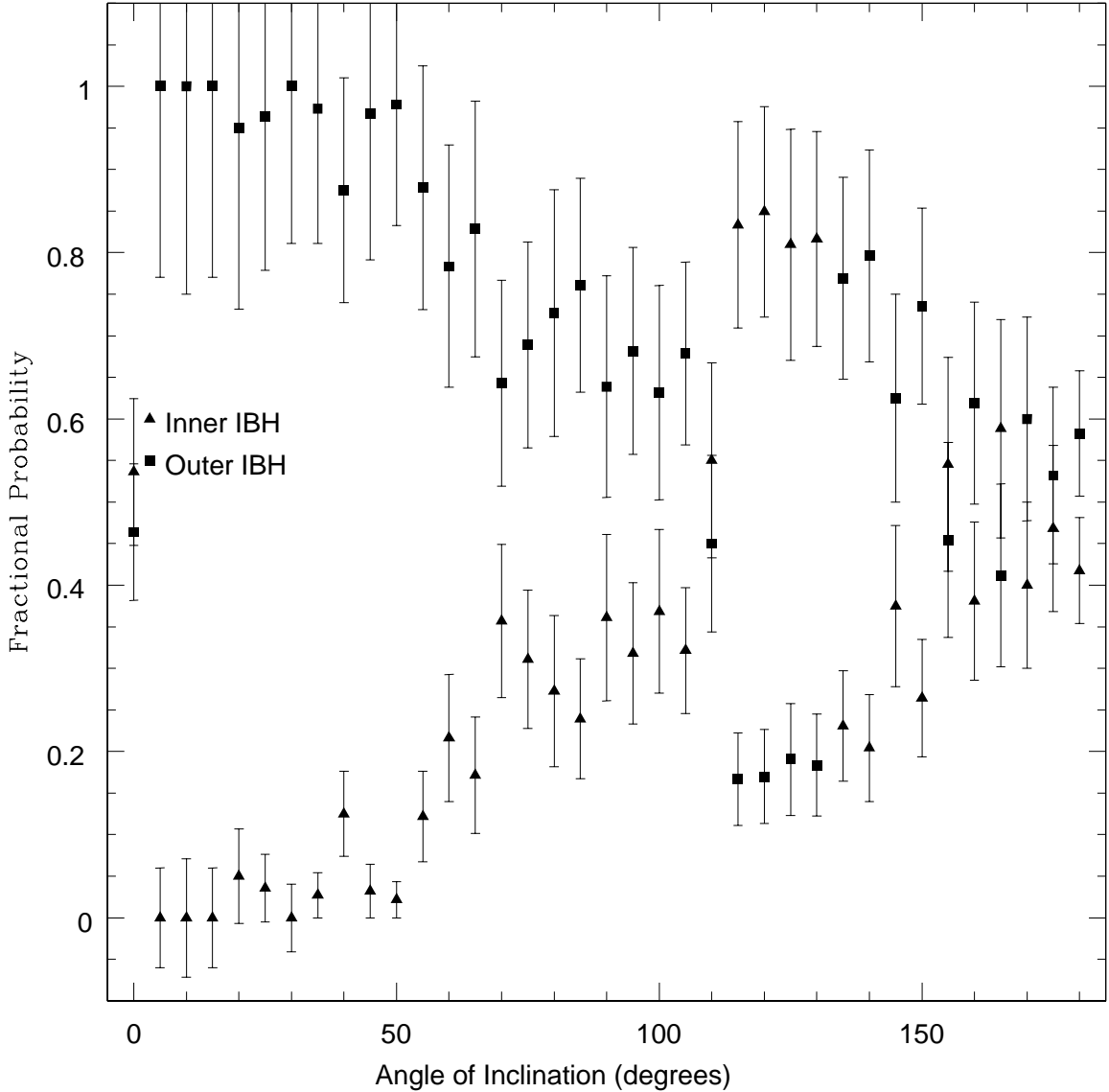}
 \caption{Probabilities with which a given IBH (either inner or outer) was
	  ejected, plotted as a function of the initial relative angle of 
	  inclination, $\Delta i_0$.  For values of $\Delta i_0$ 
	  less than $75^\o$ the outer IBH had a greater probability for
	  being ejected than the inner IBH.  As $\Delta i_0 \rightarrow 90^\o$,
	  the probability that the outer IBH decreases until the two
	  IBH have essentially an equal probability of being ejected.  As
	  the initial orbits become retrograde with respect to each other
	  the inner IBH has a slightly greater probability of being ejected
	  which slowly decreases as $\Delta i_0 \rightarrow 180^\o$
          }
 \label{fig:ejprobs}
\end{figure}

\begin{figure}
 \plotone{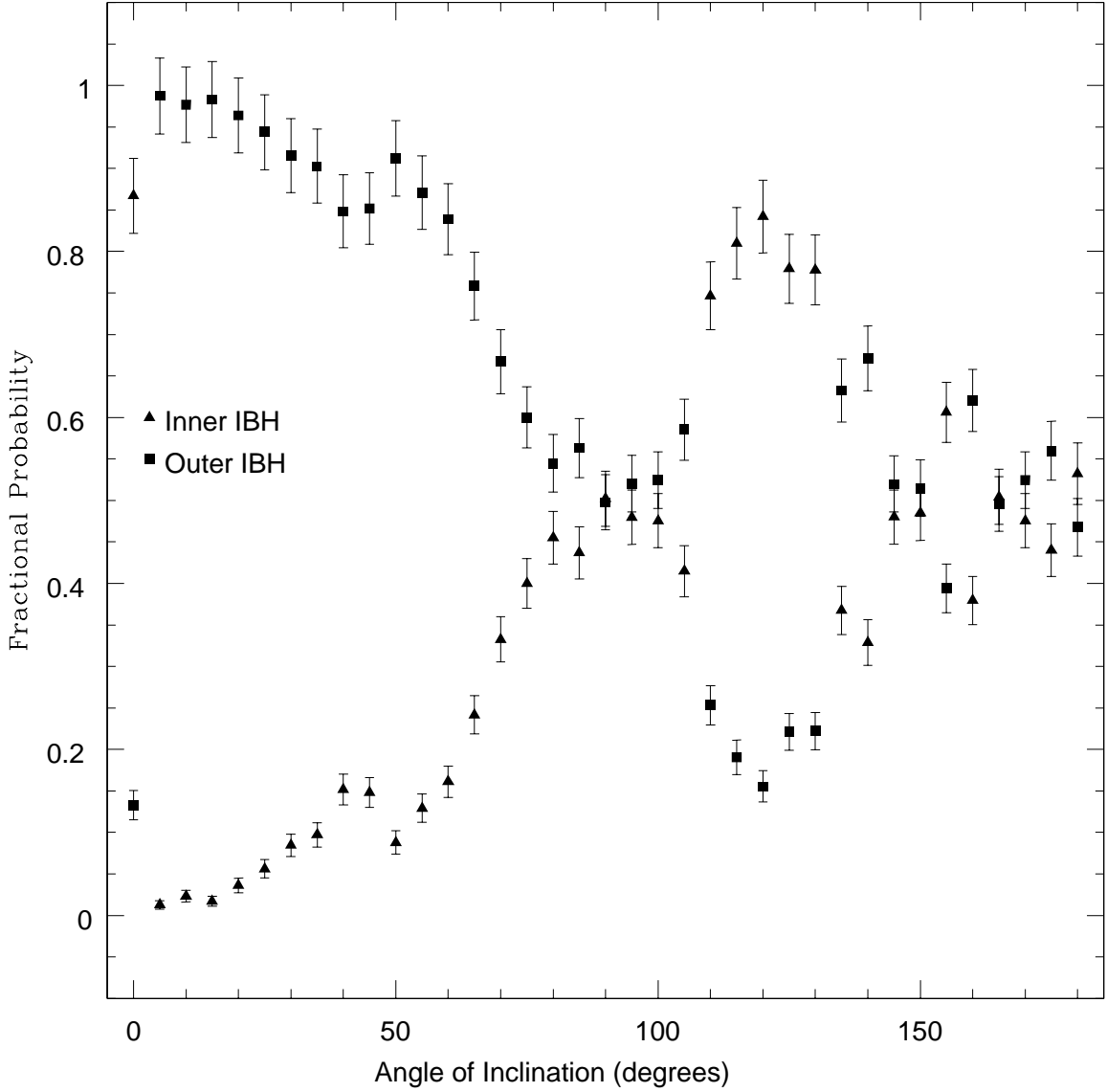}
 \caption{
	  Probabilities of which IBH (initially inner or outer) that
          merged with the CBH plotted as a function of the initial 
	  relative angle of inclination, $\Delta i_0$.  Surprisingly,
	  much like the ejection probabilities plotted in 
	  figure (\ref{fig:ejprobs}), the outer IBH is most likely to
	  merge with the CBH for $\Delta i_0 < 75^\o$.  Unlike the
	  probability with which IBH is ejected, there is a more distinct
	  transition at $\Delta i_0 = 90^\o$ where the inner IBH becomes
	  more likely to merge with the CBH.  Lastly, as
	  $\Delta i_0 \rightarrow 180^\o$ the two IBH eccentially have
	  the same probability of merging with the CBH.
          }
 \label{fig:mgprobs}
\end{figure}

\begin{figure}
 \plotone{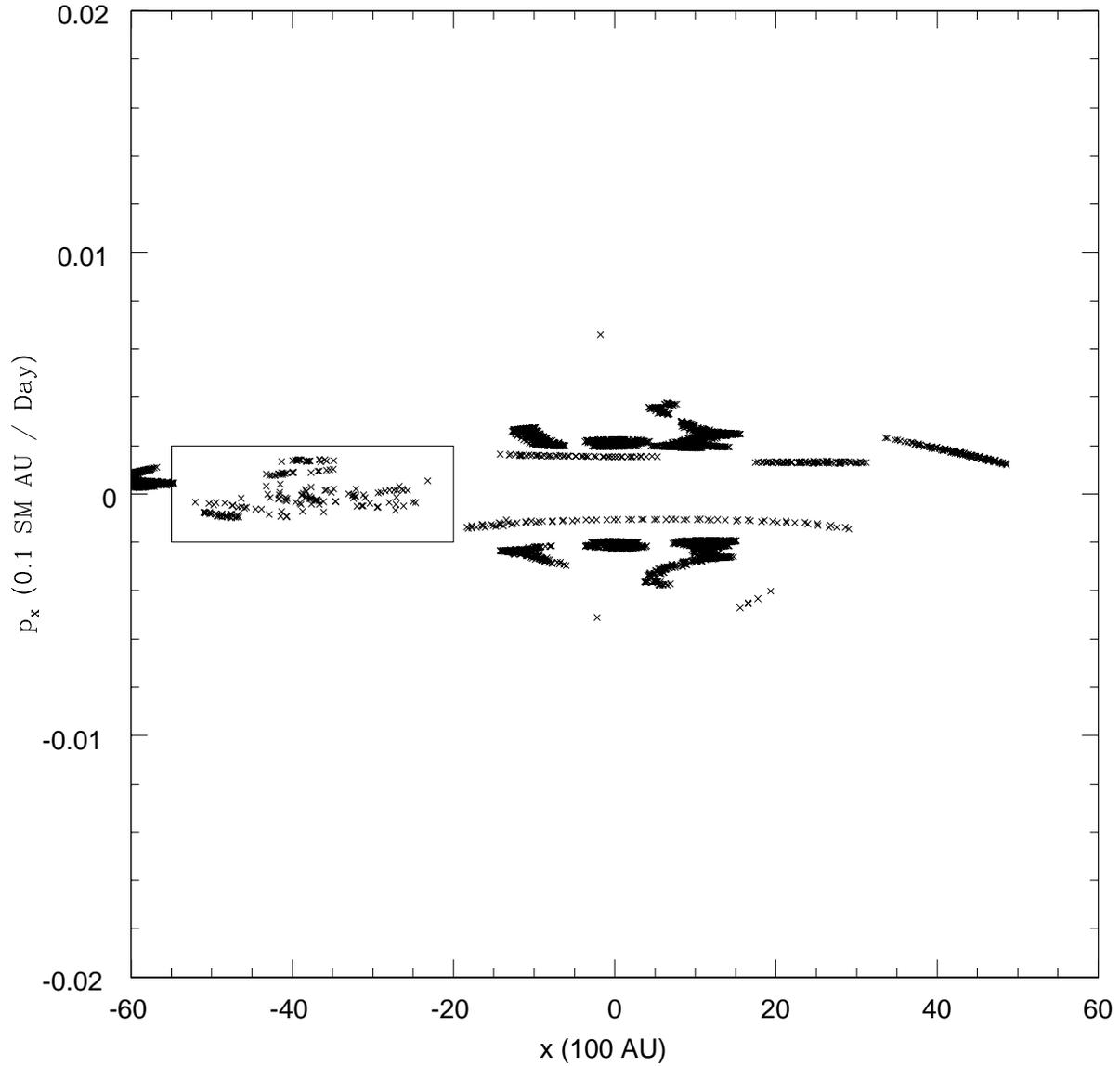}
 \caption{Phase space diagram (momentum and position for a fixed axis) for an integration resulting in the ejection of
          of the outer IBH.  Although the dynamical friction and to a lesser
          extent, general relativistic effects, "smear" the phase space plot,
          the boxed region shows behavior consistant with orbital scattering.}
 \label{fig:PhsSpceEJ}
\end{figure}

\begin{figure}
 \plotone{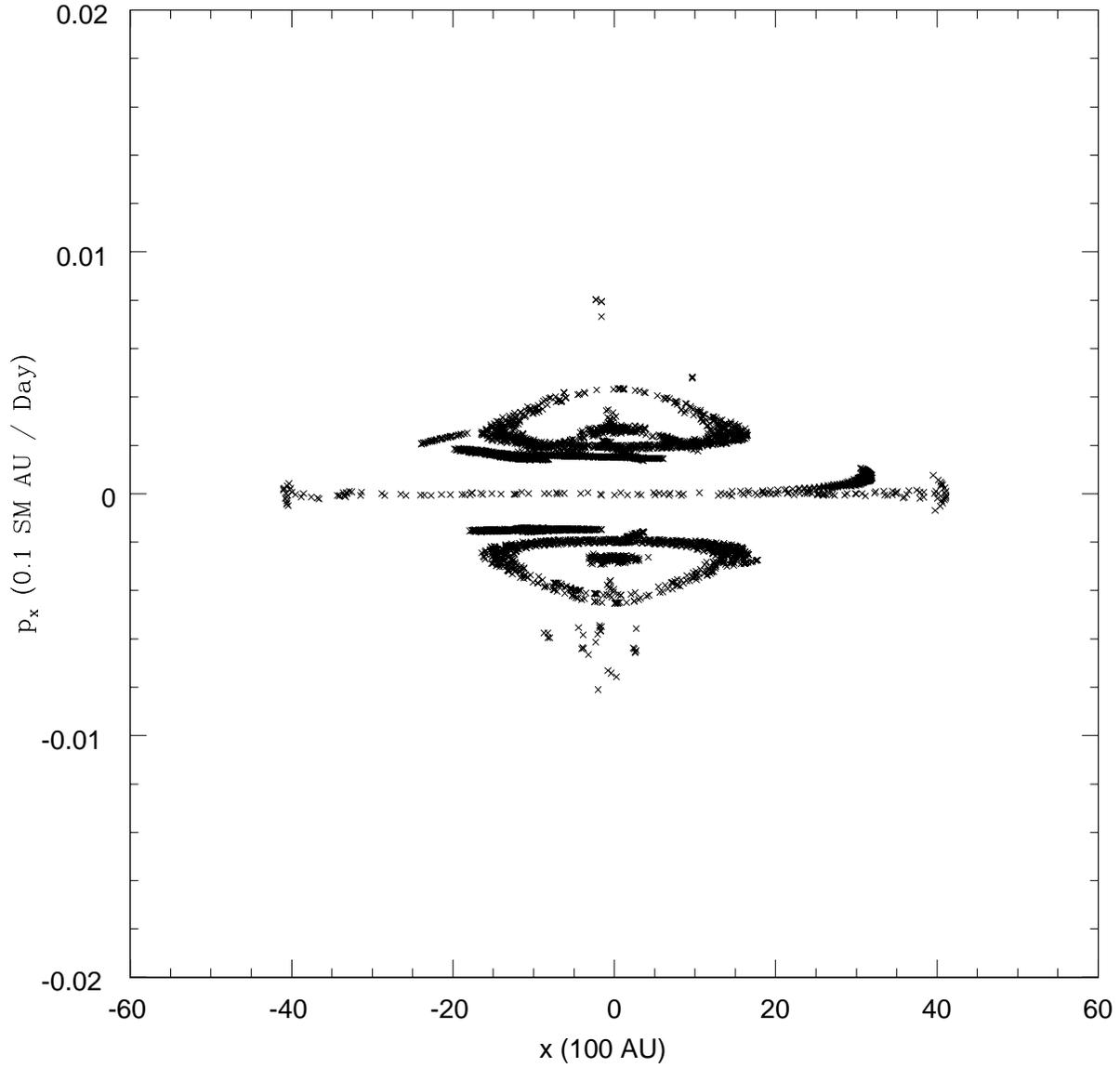}
 \caption{Phase space diagram for an integration resulting in an IBH-CBH
          merger.  Unlike figure \ref{fig:PhsSpceEJ}, the only "smearing"
          of the phase space plot is due to dynamical friction and
          relativistic effects.}
 \label{fig:PhsSpceMG}
\end{figure}

Surprisingly, the probability of an IBH-CBH merger 
had essentially no dependance on the initial angle between 
the orbital planes of the two black holes, with the
exception of the geometrically restricted special cases
 $\Delta i_0 = 0 \dgrs$ and $\Delta i_0 = 180 \dgrs$.
Figure (\ref{fig:probs}) is a plot of the probability of a merger or
ejection occurring, figure 
(\ref{fig:ejprobs}) the probability 
of a given IBH either being ejected and (\ref{fig:mgprobs}) the probability 
or merging with the CBH as a function of inclination angle.
The largest out of plane dependence of the branching ratios is associated with
$\Delta i_0 = 0 \dgrs$ and
$\Delta i_0 = 180 \dgrs$, in which case the planar geometry excludes
the oscillations in inclination that aid in eccentricity growth \citep{Iwasawa2005}.
Although not as dramatic, there appears to be a slight increase (decrease) in the
probability of an IBH being ejected from the Galactic center (merging with the CBH)
between $\Delta i_0 \sim 20 \dgrs$ and $\Delta i_0 \sim 50 \dgrs$, this could be
explained by the increased influence of the Kozai effect.
In comparing these two special cases, we find 
 it is
almost 50\% more likely for an ejection to occur for $\Delta i_0 = 180 \dgrs$ 
than it is for $\Delta i_0 = 0 \dgrs$ because an 
orbital resonance is established
for $\Delta i_0 = 0^{\o}$, and not for $\Delta i_0 = 180^{\o}$, which transfers
energy between the two IBHs in such a way that one of the IBH developes a highly
eccentric, low periastron orbit when $\Delta i_0 = 0^{\o}$.  
Furthermore, the phase space plots of two separate simulations for the same
value of $\Delta i_0$ support the belief that ejections are the result of 
IBH-IBH scattering events and an IBH-CBH merger is the result of an IBH-IBH
orbital resonance (figures \ref{fig:PhsSpceEJ} and \ref{fig:PhsSpceMG}).
The boxed region in figure (\ref{fig:PhsSpceEJ}) shows that the orbital energy
of one of the IBH had been changed abruptly giving the IBH access to a 
previously ``forbidden'' region.  Unlike the phase space diagram for the 
ejected IBH, figure (\ref{fig:PhsSpceMG}) has clearly defined regions without
any abrupt penetration into a ``forbidden'' region.

The strong mutual interactions between the IBH do not always maintain a strong
enough repulsion to prevent close encounters between the IBH. When such close
encounters occur, they alter the orbits of the IBH to the point that the resonance
interaction is broken. A sequence of close encounters follows until one of the
energy exchanges is strong enough to eject one of the IBH from the central parsec,
leaving the other bound in an eccentric orbit.
However, the three body interaction between the two IBHs and the CBH does not 
transfer enough energy to the `ejected' IBH such that it is capable of breaking 
free of the Galactic potential.  In fact, the average energy of the ejected IBH was
found to be $\sim 6 \times 10^{-11} {\rm M}_{\odot}c^2$.  Crudely accounting for dynamical
friction the IBH would then begin another infall at $\sim 5 - 10 \pc$, thereby
returning to the inner $0.05 \pc$ of the Galactic center $\sim 100 \Myr$ 
assuming a comparable $t_{fric}$ used in our calculations. 

Although the majority of all IBH-CBH interactions leave a single IBH in orbit
around the CBH, 
there appeared to be a small subset of IBH-CBH mergers, $\lesssim 1 \%$,
in which the other IBH was unbound.  Unlike the majority
of the IBH-CBH mergers (described above), these were most-likely the result of an
IBH-IBH scattering event that transfered enough energy from one of the IBHs to cause
the other to be ejected.  The IBH that remained bound was sufficiently eccentric that
it then radiated the
remainder of its orbital energy as gravitational radiation and spiraled down to merge with the CBH.
Thus, in these rare cases, it is possible to end up with neither of the initial IBH in orbit
around the CBH.

\begin{figure}
 \includegraphics[scale=0.5, angle=270]{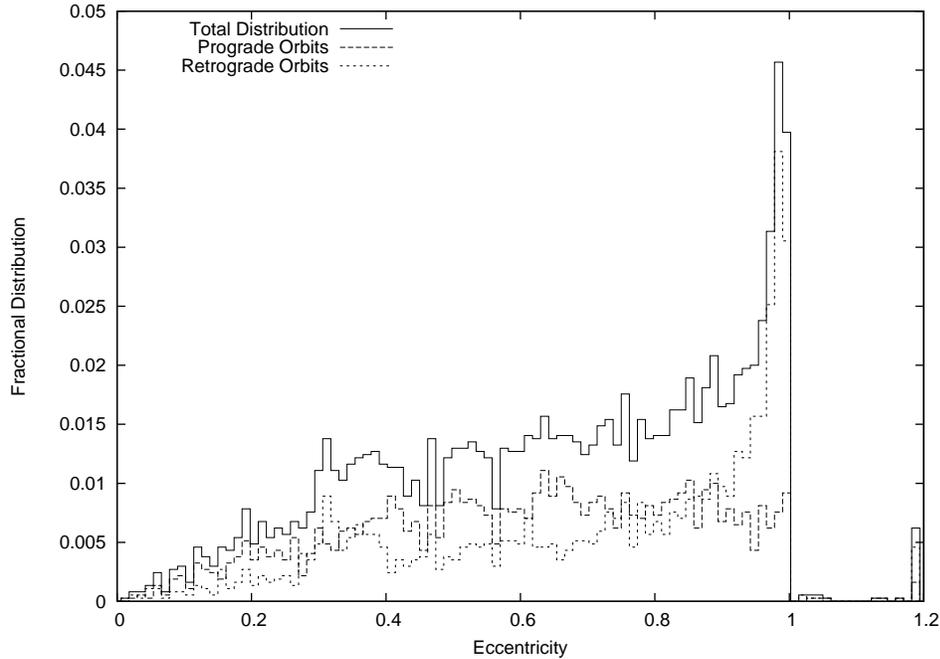}
% \plotone{EccenDist.eps}
 \caption{Distribution of the eccentricity of the remaining IBH.  Note 
	  that all eccentricities greater than 1.2 were binned as 
	  having an eccentricity of 1.2.  These are highly improbable,
	  occurring $\lesssim 0.1 \%$ of the time and are the result
          of an IBH-IBH interaction that ejected one IBH (the 
	  ``remaining'' IBH) where the other IBH lost enough energy
	  to have an orbit close enough to the CBH to radiate its remaining
	  orbital energy via gravitational radiation.
	  Additionally, notice that $\gtrsim 80 \%$ of the final
	  eccentricities greater than 0.8 are associated with retrograde
          orbits.  
	  }
 \label{fig:FinalEccen}
\end{figure}

\begin{figure}
 \plotone{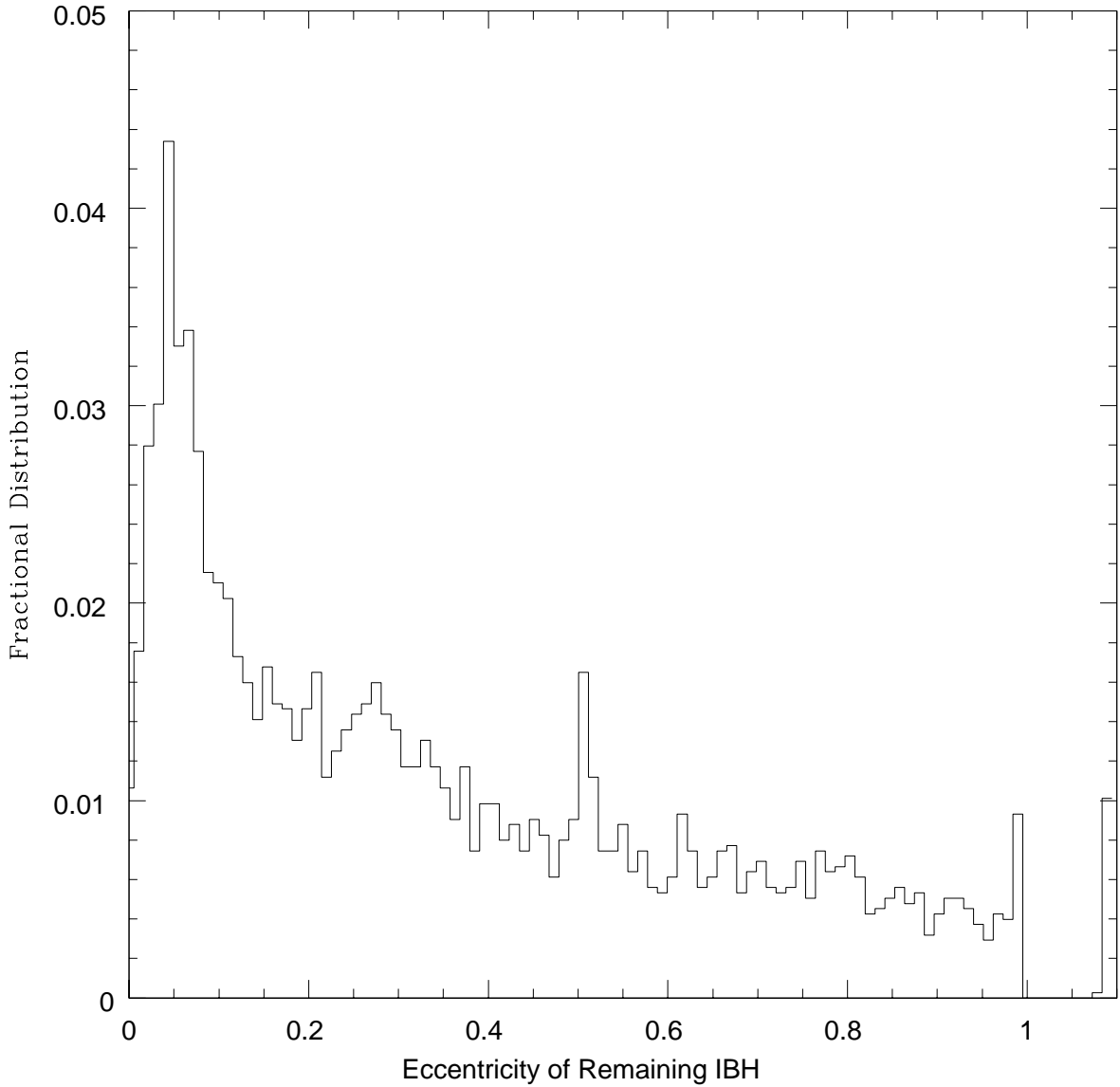}
 \caption{Distribution of the eccentricity of the remaining IBH where the 
          initial eccentricity of the inner IBH was 0.9.  Although the 
          average eccentricity of the remaining IBH is $\sim 0.46$, the
          most probable eccentricity is $\sim 0.05$ which implies that
          after a third IBH falls into the inner parsec, the system is 
          essentially back to the original configuration.  Furthermore,
          the distribution of eccentricities $\ge 1.1$ were combined and 
          are the result of an IBH-CBH merger that was caused by a 
          dramatic transfer of energy ejecting the other IBH.  Lastly,
          the probability of merger or ejection is not affected by the
          large initial eccentricity of the inner IBH.}
 \label{fig:EccenDist2}
\end{figure}

Figure (\ref{fig:FinalEccen}) is a plot of the distribution of eccentricity of
the remaining IBH, for which the average value 
 is $0.676 \pm 0.005$,
and which is skewed more towards higher eccentricities. 
Indeed, the majority of orbits have an eccentricity of $\sim 0.9$.
Additionally, we show the small fraction, $\lesssim 0.1 \%$, of cases 
where the ``remaining'' IBH had an eccentricity of greater than 1.  These
were the result of the low probability events mentioned above, in which one
IBH is ejected and the other merges with the CBH.
In order to test whether or not our initial conditions were a realistic
simulation of the system, further simulations 
of the IBH-IBH-CBH triple system were performed, in which the inner
IBH was given an initial orbit with a large eccentricity ($ \sim 0.9$).
These yielded 
 similar results as before, with
the exception that the system evolved much more quickly, taking 
$\sim 1/10$ the amount of time required when the initial eccentricity
of the inner IBH was $\lesssim 0.2$.  
Furthermore, the average eccentricity of the
remaining IBH from these simulations was $\lesssim 0.2$ with the majority
of the remaining orbits with an eccentricity $< 0.1$ (figure 
\ref{fig:EccenDist2}). This is a consequence of the fact that,
 when the inner IBH had
a highly eccentric orbit with low pericenter, it was more susceptible
to a rapid merger with the CBH, as the initial resonant interactions
drive the periastron to smaller values.
 In these cases, the inner IBH was typically
merged with the CBH prior to dramatically affecting the outer IBH's
eccentricity. The cases that resulted in ejections behaved similarly
as before, since the mutual close encounters between the IBH quickly
erased any memory of the initial conditions.
Therefore, even
though the remaining IBH after the initial IBH-IBH-CBH system 
has resulted in either an IBH-CBH merger or IBH ejection may 
have a highly eccentric orbit, a second 
IBH-IBH-CBH interaction yields branching ratios for an IBH-CBH merger
(or IBH ejection) similar to those that result when starting with a small eccentricity ($\le 0.3$) 
(within statistical error) and will mostly likely lead to initial conditions 
for the IBH remaining after this IBH-IBH-CBH interaction similar to
those of the original scenario.
Thus, overall, our simulations yield 
 branching ratios of 0.65 and 0.35
for mergers and ejections respectively (figure \ref{fig:probs}).  Coupling
this with an infall rate consistent with 
the ages of the stars discussed in the introduction, results
in an IBH-CBH merger $\lesssim 15$ Myrs, whether the 
the inner IBH has a small initial eccentricity, $\lesssim 0.3$ or a large
initial eccentricity $\gtrsim 0.7$.  This supports the idea that supermassive
black holes could be created by successive mergers of IBHs over a long 
period of time.
We also found magnitude of the gravitational radiation emitted 
during the IBH-CBH merger is consistent with \citet{Iwasawa2005} and 
should be able to be detected by LISA.

\section{Discussion}
\label{sec:Discuss}

\begin{figure}
 \plotone{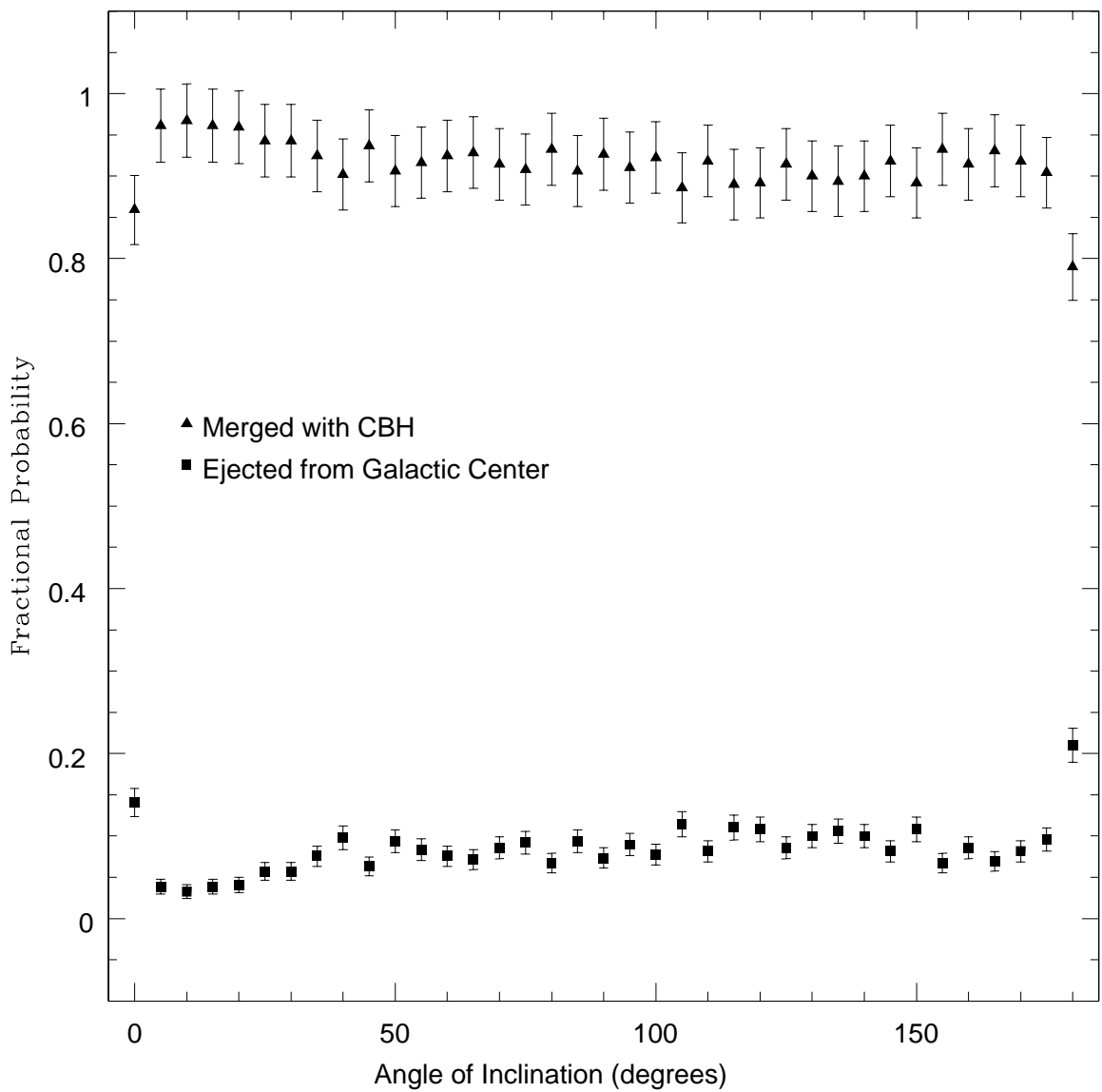}
 \caption{Branching ratios as a function of initial relative angle
          of inclination angle between the two IBH, $\Delta i_0$.
	  These probabilities correspond with the case of the inner
	  IBH initially having a circular orbit, although subsequent 
	  simulations with the initial eccentricity of the inner IBH
	  equal to 0.2, 0.3, 0.6 and 0.9 resulted in probabilities 
	  within the error of those presented above.}
 \label{fig:probs}
\end{figure}

The model problem discussed here was motivated by the desire to understand
the consequences of a steady rate of infall of IBH into the Galactic center,
a potential consequence of theories for the origins of the young stars observed
there. Perhaps surprisingly, we find that 
 the probability of a merger between the CBH and one of the IBH is 
 greater than the probability of an ejection (figure \ref{fig:probs}). 
Coupling the observed branching ratio of 0.65 for an IBH-CBH merger,
along with the infall rate $\sim 10^7$ yrs 
(consistent with the ages of  the young stars found in the central parsec),
 results in an IBH-CBH merger occurring approximately once every 11 Myrs.
We have also found similar results for both enhanced and reduced dynamical
friction ($t_{fric} \sim 10^6$ yrs and $t_{fric} \sim 10^8$ yrs)
where the differences are associated with statistical error and
the total time for infall and IBH-IBH interactions to occur.
Depending on the rate of IBH infall, such mergers may have contributed significantly to the mass growth of the
CBH \citep{Ebisuzaki01}.  

The eccentricity of a surviving IBH is determined by several factors. 
Dynamical friction will circularise the orbit (e.g. \citet{Berukoff06}),
although this depends on the background density profile. Indeed, eccentricity
can be pumped if there is an evacuated core, so that dynamical friction 
operates over only part of the orbit, as we have seen in this calculation and
has been verified in numerical simulations such as \citet{LockBaum}. In our
case, these effects are superceded by the strong interactions between the pair
of IBH, both in the form of resonant interactions and close passages and scattering. 
However, we find that a system that starts with an inner IBH on an eccentric orbit
is likely to leave a survivor on a low eccentricity orbit, and vice versa. 
Taken together, this implies that the outcome of a series of IBH infalls can be
considered as a chain of three-body encounters such as the ones described here,
unless the rate of infall is considerably faster than one every $10^7$~years.

In conclusion, although the stellar scattering responsible for
dynamical friction cannot cause an IBH-CBH merger in a Hubble time,
the dynamical interactions between two or more IBH result in ejection
or merger on timescales $\sim 10^7$~years, so that the number of IBH
inhabiting the central parsec today should be small in number, even
if the process of infall is a regular feature of Galactic center life.

\acknowledgements
The authors acknowledge useful conversations with Steve Berukoff regarding
his work simulating the scenario feeding our initial conditions and 
John Chambers regarding the operation of {\em Mercury6}.

\bibliographystyle{apj}
\bibliography{ms}

\clearpage

%\begin{figure}
% \plotone{edist.eps}
% \caption{Histogram of the eccentricities for the remaining IBH}
% \label{fig:edist}
%\end{figure}

\end{document}